\documentclass[12pt]{article}

\usepackage[utf8]{inputenc}
\usepackage[T1]{fontenc}
\usepackage{geometry}
\usepackage{graphicx}
\usepackage{amsmath,amssymb}
\usepackage{booktabs}
\usepackage{hyperref}
\usepackage[numbers,sort&compress]{natbib}
\usepackage{xcolor}
\usepackage{multirow}
\usepackage{longtable}
\usepackage{float}
\usepackage{setspace}
\usepackage{url}
\usepackage[title]{appendix}
\usepackage{tabularx}
\usepackage{array}
\usepackage{caption}

\hypersetup{
  colorlinks=true,
  linkcolor=blue,
  citecolor=blue,
  urlcolor=blue
}

\date{}

\begin{document}

\begin{center}
{\LARGE\bfseries Depictions of Depression in Generative AI Video Models:\\ A Preliminary Study of OpenAI's Sora 2\par}
\vspace{12pt}
{\large
Matthew Flathers\textsuperscript{1},
Griffin Smith\textsuperscript{2},
Julian Herpertz\textsuperscript{1,3},
Zhitong Zhou\textsuperscript{1,4},
John Torous\textsuperscript{5}\par}
\vspace{8pt}
{\small
\textsuperscript{1}Division of Digital Psychiatry, Beth Israel Deaconess Medical Center, Boston, MA, USA\\
\textsuperscript{2}Rhode Island School of Design, Providence, RI, USA\\
\textsuperscript{3}Department of Psychiatry and Neuroscience, Charit\'{e} Berlin University Medicine,\\
Campus Benjamin Franklin, Berlin, Germany\\
\textsuperscript{4}Boston University, School of Public Health, Boston, MA, USA\\
\textsuperscript{5}Department of Psychiatry, Beth Israel Deaconess Medical Center, Boston, MA, USA\par}
\end{center}
\vspace{12pt}

\begin{abstract}
\noindent\textbf{Background:} Generative video models are increasingly capable of producing complex depictions of mental health experiences, yet little is known about how these systems represent conditions like depression. Because AI-generated content may reach people during vulnerable periods, understanding what visual narratives these models produce for sensitive concepts carries clinical relevance.

\noindent\textbf{Objective:} This study aimed to characterize how OpenAI's Sora 2 generative video model depicts depression, and to examine whether depictions differ between the consumer App and developer API access points, which differ in their product-layer mediation.

\noindent\textbf{Methods:} We generated 100 videos using the single-word prompt ``Depression'' across two access points: the consumer App (n=50) and developer API (n=50). Two trained coders independently coded narrative structure, visual environments, objects, figure demographics, and figure states. Inter-rater reliability was assessed using Cohen's kappa, with dimensions showing insufficient agreement excluded from analysis. Computational features (visual aesthetics, audio, semantic content, temporal dynamics) were extracted and compared between modalities using Welch's t-tests with Benjamini-Hochberg false discovery rate correction.

\noindent\textbf{Results:} App-generated videos exhibited a pronounced recovery bias: 78\% (39/50) featured narrative arcs progressing from depressive states toward resolution, compared with 14\% (7/50) of API outputs. This divergence was reinforced across channels. App videos brightened over time (slope = 2.90 brightness units/second vs. -0.18 for API; d = 1.59, q < .001) and contained three times more motion (d = 2.07, q < .001). Across both modalities, videos converged on a narrow visual vocabulary: predominantly seated figures (93\%), downward gaze (96\%), and recurring objects including hoodies (n=194), windows (n=148), and rain (n=83). Transcript language in depressive phases emphasized weight and containment (``heavy,'' ``drowning,'' ``room''), while recovery phases reversed these patterns: brightness increased 27\% (d = 0.70, p < .001), gaze shifted upward in 68\% of recovery videos, and terms like ``light'' and ``breath'' emerged. Figures were predominantly young adults (88\% aged 20-30) and nearly always alone (98\%). Gender varied by access point: App outputs skewed male (68\%), API outputs skewed female (59\%).

\noindent\textbf{Conclusions:} Sora 2 does not invent new visual grammars for depression but compresses and recombines cultural iconographies, while platform-level constraints substantially shape which narratives reach users. Clinicians should be aware that AI-generated mental health video content reflects training data and platform design rather than clinical knowledge, and that patients may encounter such content during vulnerable periods.

\end{abstract}


\section{Introduction}

Media representations of psychiatric conditions shape how the public understands mental illness, and how patients understand their own diagnoses \cite{ref1,ref2,ref3,ref4}. Increasingly, these representations are circulating through short-form video (SFV) platforms. Mental health has emerged as one of the most prevalent topics on TikTok, Instagram Reels, and YouTube Shorts, with hashtags like \#mentalhealth accumulating over 44 billion views on TikTok alone \cite{ref5}. SFV platforms have become embedded in daily life, with roughly six-in-ten U.S. teens visiting TikTok daily and one-third using at least one major platform almost constantly \cite{ref6,ref7}.

Against this backdrop, generative artificial intelligence has introduced new capabilities for creating custom SFV content at scale. AI-generated videos can now be produced much faster and more cheaply than traditional video editing pipelines. OpenAI's Sora 2, released in September 2025, is an early leader in this space \cite{ref8}. Within 72 hours of launch, the Sora mobile application became the top-downloaded app on the iOS App Store, surpassing ChatGPT with over 164,000 downloads \cite{ref9}. The app functions as a social network for AI-generated videos, featuring a vertical feed of 10-second clips with likes, comments, and remix tools \cite{ref8}. In December 2025, OpenAI announced a partnership with Disney that will bring over 200 characters from Disney, Marvel, Pixar, and Star Wars to the platform \cite{ref10}. OpenAI's policies prohibit use by children under 13 \cite{ref11}, licensing characters from franchises with young audiences creates obvious pressure on these boundaries. As these capabilities expand beyond entertainment, AI videos will increasingly reach people who are seeking to understand their own mental health.

Sora 2 is not alone. Multiple labs have released generative AI video models that can produce highly realistic outputs that viewers cannot always reliably distinguish from authentic footage \cite{ref12}. Google's Veo 3 pairs high-fidelity video generation with native audio generation \cite{ref13}, Runway's Gen-4.5 provides sophisticated video editing capabilities \cite{ref14}, Meta has launched Vibes to integrate AI video generation into its social ecosystem \cite{ref15}, and ByteDance’s Seedance demonstrates high-fidelity multi-shot video generation with consistent subject representation across scenes \cite{ref16}.  Most generative video tools build upon diffusion image model architectures, much like AI Image models. Unlike transformer-based language models that predict and generate words sequentially, diffusion models learn to transform noise into coherent imagery. They do this through a two-phase process: systematically adding noise to images until they become visual static (forward diffusion), then learning to reverse this process to reconstruct coherent images \cite{ref17,ref18}. Through training on billions of examples, these models learn the contours, patterns, and statistical regularities of their training data, storing this information across vast parameter spaces. Video diffusion models extend this into higher-dimensional space, maintaining consistency across hundreds of frames while generating plausible motion and transitions \cite{ref19,ref20}.

The outputs of these systems occupy a unique categorical space \cite{ref21}. The underlying models train on heterogeneous sources simultaneously: clinical documentation, Hollywood films, pharmaceutical advertising, video blogs and other user-generated content. The resulting synthesis reveals aggregate cultural understanding rather than any single authoritative source. When a user prompts a model with ``Depression,'' the resulting artifact is not summoned from nothing, nor is it copied from some original human source. The user's prompt is mediated by a technical apparatus that constrains what can be produced to what its programming makes possible \cite{ref22}. The output is ``found'' by the AI model in the diffusion process, selected from the space of possible outputs that training made probable. Whether the patterns learned during training extend beyond what things look like to capture implicit associations, co-occurrences, and cultural conventions embedded in the training data (as has been observed in image \cite{ref23}, text \cite{ref24}, and embedding \cite{ref25} models) remains an open question this study begins to address.

This study examines the Sora 2 generative AI video system, but our methodological approach addresses a broader issue in clinical AI research. Much existing work treats ``the model'' as the object of study without distinguishing between 1) the underlying models and 2) the products and consumer-facing apps/tools built upon them. When a user submits a prompt into the Sora app, they interact with a system that includes the model plus product layers. When a researcher accesses the same model through the API, fewer of these layers intervene (See Figure 1 for a generalized illustration of this distinction). Patterns observed in an app may therefore reflect product-level decisions (e.g. more safety features and filters) rather than model-level associations, or vice versa. Studying only the product tells us what users encounter but not what the model has learned; studying only the model tells us what the training has encoded but not what reaches users \cite{ref26}. Comparing both access points reveals something about the safety and content moderation layers that companies have implemented for sensitive topics like depression. The distinction between model and product is also important for research on safety evaluation within the wider AI product ecosystem. Future developers will build applications on top of AI models accessed via APIs, not the consumer product layer. Sora 2 provides an opportunity to examine these dynamics because OpenAI has made both access points available: the consumer app and the developer API.

\begin{figure}[htbp]
\centering
\includegraphics[width=\textwidth]{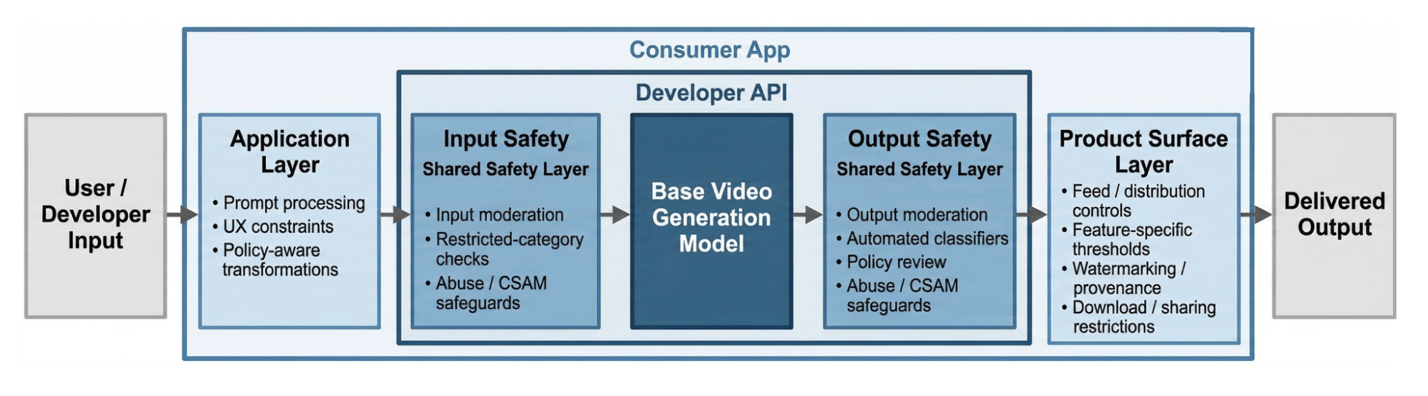}
\caption{Generalized system architecture for generative video AI platforms, illustrating the distinction between Developer API and Consumer App access pathways. Both pathways share core safety infrastructure (center), including input and output moderation. Consumer Apps additionally route prompts through an application layer (prompt processing, UX constraints, policy-aware transformations) before generation, and through a product surface layer (feed curation, watermarking, distribution controls) after. The Developer API bypasses these layers, accessing only the shared safety infrastructure and base model.}
\label{fig:architecture}
\end{figure}

When someone experiencing a depressive episode searches for content about depression and is served AI-generated videos on platforms like Sora 2, those outputs become part of their informational environment during a vulnerable period. The content might reinforce their experience, normalize or stigmatize it, or offer implicit models of recovery or self-harm \cite{ref4,ref27,ref28}. Whether AI-generated depictions of depression draw on clinical understanding, popular media tropes, pharmaceutical advertising or some mixture remains an empirical question. The answer carries implications for how clinicians think about the information environment surrounding their patients.

\section{Methods}

\subsection{Data Generation}

We generated 50 videos using each of two access points to OpenAI's Sora 2 system: the consumer-facing mobile app and the developer API, yielding a total corpus of 100 videos. All videos were generated in portrait orientation (9:16 aspect ratio) with each platform's standard output parameters: 10-second clips at 720p resolution from the app and 12-second clips at 720p resolution using the Sora 2 endpoint (sora-2) via the OpenAI Responses API. To control for potential model updates, all generations occurred within a single one-week window (November 21--28, 2025). All app videos were generated from a new Sora account with no prior video generation history.

Each video was generated using the isolated prompt ``Depression'' with no additional qualifiers, contextual framing, or prompt engineering. Our minimal prompt design was a deliberate methodological choice. The prompt “Depression” is inherently ambiguous. It might refer to a clinical mood disorder, a colloquial feeling, a topological feature, or an economic phenomenon. Elaborate prompts resolve this ambiguity for the model; minimal prompts force the model to resolve it from its encoded associations. A prompt like ``a depressed person sitting alone in a dark room crying'' tests whether a model can render a specified scene but reveals little about what the model associates with depression as a concept, because the user has already supplied the interpretive content. Single-term prompts shift this interpretive labor to the model. When given only ``Depression'' with no contextual elaboration, the model must supply setting, figures, mood, color, motion, and narrative entirely from its encoded associations. Every visual choice in the output therefore reflects training data or product-level steering rather than user instruction, making minimal prompting a more sensitive probe of the model's default representations.

\subsection{Qualitative Analysis}

Two authors (MF and ZZ) independently conducted a qualitative content analysis of all 100 videos using a directed approach, in which initial coding categories were derived from prior literature and the study's research questions \cite{ref29}. Each coder tagged videos across six dimensions: (1) narrative arc (presence and direction of emotional shifts), (2) environments (physical settings with timestamps), (3) objects (visual symbols and props with timestamps), (4) figure demographics (gender, race/ethnicity, age), (5) figure states (posture, facial expression,  alone in frame with timestamps), and (6) overall notes (AI artifacts and uncertainties). Raw codes were harmonized to standardized categories prior to analysis. Extended coding procedures and dimension rationale appear in Appendix A; the full codebook appears in Appendix B; and harmonization dictionaries appear in Appendix C.

\subsubsection{Reliability}

Both authors independently coded the full dataset. One author's coding (MF) was designated as the primary dataset for analysis; the second author's (ZZ) served as inter-rater reliability verification. Cohen's kappa ($\kappa$) was calculated for presence/absence of categorical elements, and for dimensions involving temporal segments, we also calculated intersection-over-union (IoU) at one-second resolution to assess agreement on timing. Given the scale of the dataset, item-by-item consensus resolution was not feasible; reliability metrics verified that the coding scheme could be applied consistently across coders.

\subsection{Quantitative Analysis}

To complement qualitative coding, we extracted computational features across four domains: visual aesthetics, audio properties, semantic content, and temporal dynamics. Visual features included brightness, saturation, colorfulness, color temperature, and GLCM texture measures \cite{ref30}, sampled at one-second intervals. Audio analysis extracted volume, pitch, and spectral centroid using the librosa library \cite{ref31}. Semantic analysis of transcribed speech (OpenAI Whisper \cite{ref32}) included VADER sentiment \cite{ref33}, custom light/dark lexicons based on conceptual metaphor theory \cite{ref34}, and linguistic markers including first-person pronoun ratio and negation frequency. Temporal dynamics were captured through dense optical flow (Farneb{\"a}ck algorithm \cite{ref35}), scene detection, and per-second feature trajectories. Features were selected based on documented connections to depression representation in prior research; technical specifications and literature rationale appear in Appendix D.

\subsection{Ethical Considerations}

This study analyzed AI-generated video content produced by a commercially available system (OpenAI Sora 2) and did not involve human participants, human biological materials, or identifiable personal data. No human subjects were recruited, surveyed, or observed, and no personally identifiable information was collected or analyzed. Accordingly, institutional review board review was not required under federal regulations (45 CFR 46) or institutional policy. All analyzed content was generated by the research team using publicly available AI tools.

\subsection{Statistical Analysis}

For each quantitative feature, we computed the mean and standard deviation for each access modality. Differences between app and API outputs were assessed using Welch's t-tests \cite{ref36}, with Benjamini-Hochberg false discovery rate correction \cite{ref37} to control for multiple comparisons across all 28 features tested. Effect sizes were quantified using Cohen's d \cite{ref38}. To characterize temporal dynamics, we computed the linear slope of each feature's time series and compared slope distributions between modalities using the same inferential approach.

\section{Results}

\subsection{Quantitative Findings}

We generated 50 videos from each access point (App and API), yielding a corpus of 100 videos for analysis. All comparisons used Welch's t-tests with Benjamini-Hochberg false discovery rate (FDR) correction; we report FDR-corrected q-values and Cohen's d effect sizes throughout.

\begin{table}[htbp]
\centering
\caption{Aggregate descriptive statistics. Mean $\pm$ SD for all visual, texture, motion, audio, speech, sentiment, and semantic features across App (n = 50) and API (n = 50) generated videos. Cohen's d effect sizes and FDR-corrected q-values from Welch's t-tests with Benjamini-Hochberg correction. * indicates q < .05.}
\label{tab:aggregate}
\footnotesize
\begin{tabular}{llcccc}
\toprule
\textbf{Category} & \textbf{Feature} & \textbf{App (M $\pm$ SD)} & \textbf{API (M $\pm$ SD)} & \textbf{\textit{d}} & \textbf{\textit{q}} \\
\midrule
\multirow{5}{*}{Visual} & Brightness Mean & 64.59 $\pm$ 13.35 & 53.48 $\pm$ 11.86 & 0.87 & 0.000* \\
 & Saturation Mean & 99.24 $\pm$ 26.61 & 92.48 $\pm$ 23.08 & 0.27 & 0.342 \\
 & Colorfulness Mean & 25.67 $\pm$ 7.48 & 22.41 $\pm$ 8.01 & 0.42 & 0.086 \\
 & Color Temperature Mean & $-$13.38 $\pm$ 10.56 & $-$14.45 $\pm$ 9.66 & 0.10 & 0.716 \\
 & Edge Density Mean & 0.03 $\pm$ 0.01 & 0.03 $\pm$ 0.01 & $-$0.31 & 0.239 \\
\midrule
\multirow{6}{*}{Texture} & GLCM Contrast & 12.15 $\pm$ 5.77 & 7.78 $\pm$ 5.11 & 0.79 & 0.001* \\
 & GLCM Dissimilarity & 1.20 $\pm$ 0.33 & 1.12 $\pm$ 0.31 & 0.22 & 0.453 \\
 & GLCM Homogeneity & 0.72 $\pm$ 0.05 & 0.70 $\pm$ 0.05 & 0.51 & 0.032* \\
 & GLCM Energy & 0.17 $\pm$ 0.04 & 0.17 $\pm$ 0.04 & $-$0.03 & 0.908 \\
 & GLCM Correlation & 0.96 $\pm$ 0.02 & 0.97 $\pm$ 0.02 & $-$0.53 & 0.025* \\
 & GLCM Entropy & 6.98 $\pm$ 0.53 & 6.94 $\pm$ 0.51 & 0.09 & 0.730 \\
\midrule
\multirow{2}{*}{Motion} & Optical Flow Mean & 0.35 $\pm$ 0.15 & 0.11 $\pm$ 0.07 & 2.07 & 0.000* \\
 & Scene Cut Count & 3.10 $\pm$ 1.88 & 0.76 $\pm$ 1.42 & 1.39 & 0.000* \\
\midrule
\multirow{3}{*}{Audio} & Audio Volume Mean & 0.10 $\pm$ 0.01 & 0.09 $\pm$ 0.01 & 0.83 & 0.000* \\
 & Audio Pitch Mean & 115.51 $\pm$ 30.04 & 148.87 $\pm$ 40.78 & $-$0.92 & 0.000* \\
 & Spectral Centroid Mean & 1744.08 $\pm$ 395.02 & 1792.82 $\pm$ 284.42 & $-$0.14 & 0.633 \\
\midrule
\multirow{2}{*}{Speech} & Word Count & 31.40 $\pm$ 5.70 & 36.50 $\pm$ 8.01 & $-$0.73 & 0.001* \\
 & Speech Rate (words/sec) & 4.18 $\pm$ 0.64 & 4.07 $\pm$ 0.54 & 0.19 & 0.552 \\
\midrule
\multirow{4}{*}{Sentiment} & Sentiment Compound & 0.19 $\pm$ 0.33 & 0.24 $\pm$ 0.36 & $-$0.15 & 0.627 \\
 & Sentiment Positive & 0.09 $\pm$ 0.05 & 0.10 $\pm$ 0.07 & $-$0.17 & 0.554 \\
 & Sentiment Negative & 0.04 $\pm$ 0.05 & 0.04 $\pm$ 0.04 & $-$0.07 & 0.780 \\
 & Sentiment Neutral & 0.87 $\pm$ 0.06 & 0.86 $\pm$ 0.08 & 0.18 & 0.552 \\
\midrule
\multirow{6}{*}{Semantic} & Light Word Count & 0.02 $\pm$ 0.14 & 0.16 $\pm$ 0.37 & $-$0.50 & 0.035* \\
 & Dark Word Count & 0.58 $\pm$ 0.67 & 0.64 $\pm$ 0.52 & $-$0.10 & 0.716 \\
 & Light/Dark Word Ratio & 0.27 $\pm$ 0.27 & 0.27 $\pm$ 0.29 & 0.00 & 1.000 \\
 & First Person Pronoun Ratio & 0.04 $\pm$ 0.04 & 0.08 $\pm$ 0.05 & $-$0.78 & 0.001* \\
 & Negation Ratio & 0.02 $\pm$ 0.03 & 0.03 $\pm$ 0.03 & $-$0.12 & 0.701 \\
 & Lexical Diversity & 0.90 $\pm$ 0.05 & 0.89 $\pm$ 0.05 & 0.22 & 0.453 \\
\bottomrule
\end{tabular}
\end{table}

\subsubsection{Visual Aesthetics}

App-generated videos were significantly brighter than API videos (M = 64.59 $\pm$ 13.35 vs. 53.48 $\pm$ 11.86; d = 0.87, q < .001). This difference emerged over time: App videos exhibited a pronounced brightening trajectory (slope = 2.90 $\pm$ 2.43 brightness units/second) while API videos remained essentially flat (slope = $-$0.18 $\pm$ 1.24; d = 1.59, q < .001). Figure 2 illustrates this divergence, showing App videos beginning at comparable brightness levels to API videos but progressively lightening across their duration.

\begin{figure}[htbp]
\centering
\includegraphics[width=\textwidth]{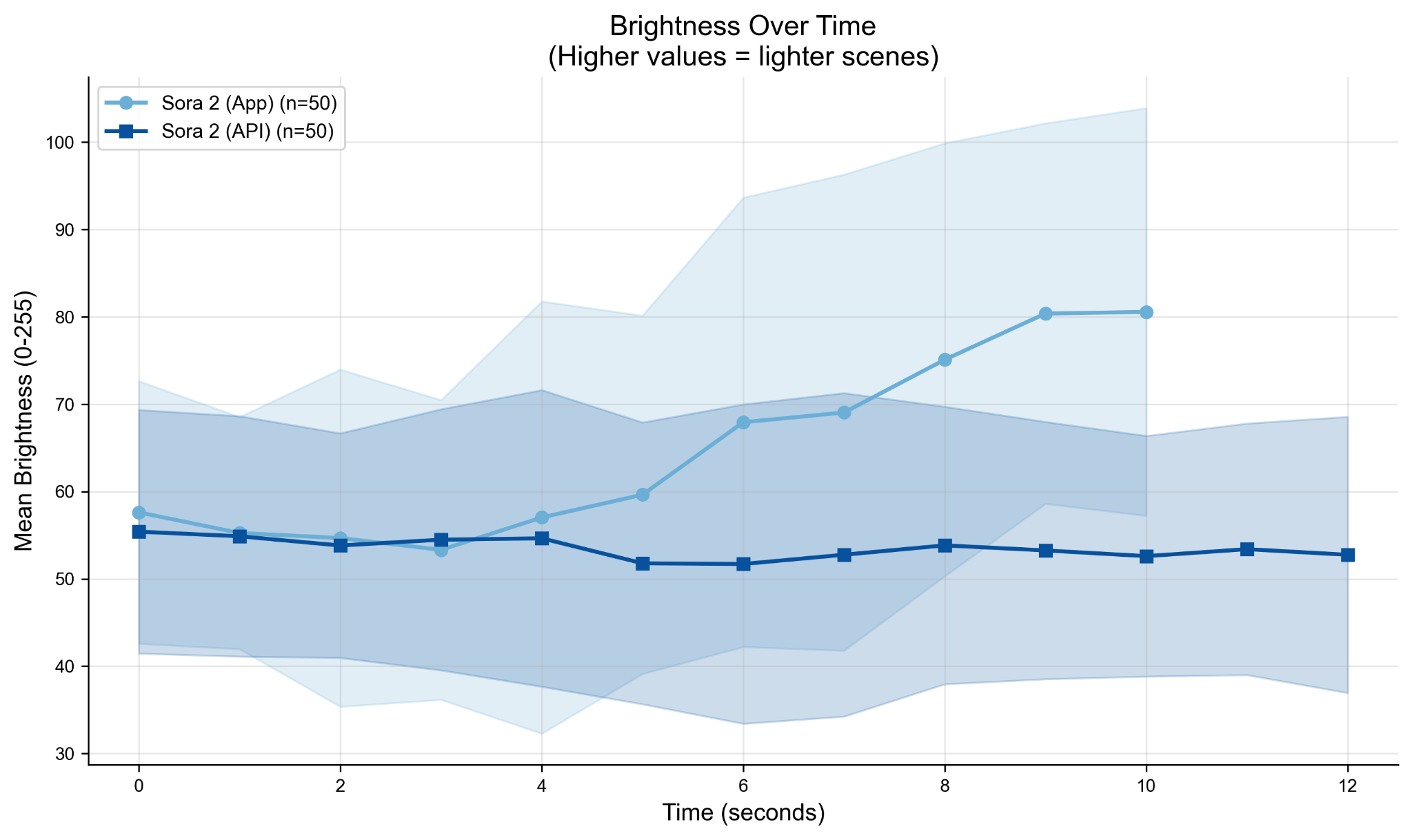}
\caption{Brightness trajectory over time. Mean brightness (0--255 scale) at each second for App (light blue, n = 50) and API (dark blue, n = 50) generated videos. Shaded regions indicate $\pm$1 SD. App videos exhibit a pronounced brightening trajectory (slope = 2.90 units/second), while API videos remain relatively flat (slope = $-$0.18). The difference in brightness slope was statistically significant (d = 1.59, q < .001, Welch's t-test with BH FDR correction).}
\label{fig:brightness}
\end{figure}

Chromatic properties showed complementary patterns. Although mean saturation did not differ significantly between modalities (q = .34), saturation trajectories diverged substantially: App videos desaturated over time (slope = $-$2.99 $\pm$ 3.17), while API videos maintained stable saturation (slope = 0.24 $\pm$ 1.30; d = $-$1.33, q < .001). Similarly, App videos warmed in color temperature over time (slope = 2.24 $\pm$ 2.26) compared to minimal change in API videos (slope = 0.18 $\pm$ 0.85; d = 1.21, q < .001). Together, these trajectories describe a characteristic App visual arc: beginning in cool, saturated tones and transitioning toward warm, desaturated, brighter imagery.

Texture analysis revealed limited additional differences. App videos exhibited higher GLCM contrast (M = 12.15 $\pm$ 5.77 vs. 7.78 $\pm$ 5.11; d = 0.79, q < .001) and homogeneity (d = 0.51, q = .03), indicating sharper local intensity variation and more uniform textural regions. Other texture features remained largely consistent across modalities.

\subsubsection{Temporal Dynamics}

The most pronounced differences emerged in motion characteristics. App videos contained approximately three times more optical flow than API videos (M = 0.35 $\pm$ 0.15 vs. 0.11 $\pm$ 0.07 pixels/frame; d = 2.07, q < .001), indicating substantially greater movement within scenes (Figure 3a). This difference persisted across the full video duration, with App videos maintaining elevated motion throughout rather than concentrating movement in particular segments.

Scene structure differed dramatically between modalities. App videos contained a mean of 3.10 $\pm$ 1.88 scene cuts, while API videos contained only 0.76 $\pm$ 1.44 cuts (d = 1.39, q < .001). Figure 3b visualizes this difference as cumulative scene accumulation over time: by the end of App videos, viewers had encountered approximately 4 distinct scenes on average, compared to fewer than 2 for API videos.

\begin{figure}[htbp]
\centering
\includegraphics[width=\textwidth]{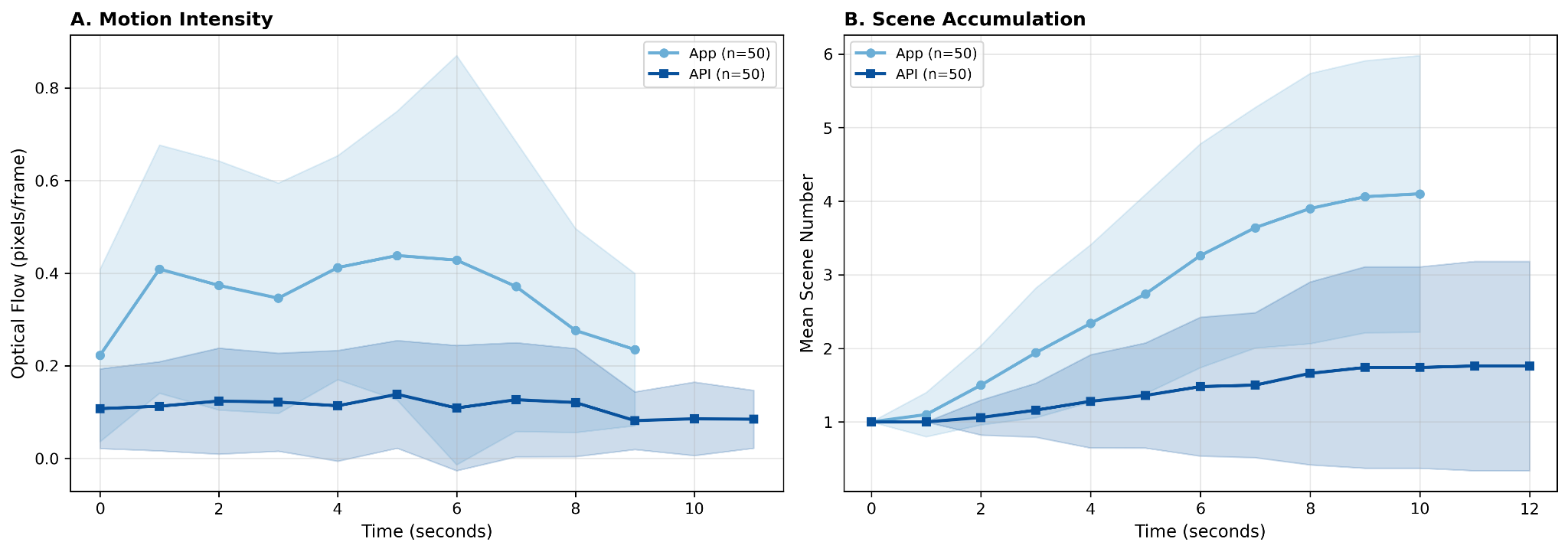}
\caption{Temporal dynamics of motion and editing in App vs. API videos. (A) Motion intensity: Mean optical flow magnitude (pixels/frame; Farneb{\"a}ck dense optical flow) at each second for App (light blue; n = 50) and API (dark blue; n = 50) videos. Shaded bands indicate $\pm$1 SD. App videos exhibit substantially higher motion across the clip duration (overall M = 0.35 vs. 0.11; d = 2.07, q < .001), indicating greater within-scene dynamism than the relatively static API outputs. (B) Scene accumulation: Mean cumulative scene number at each second, where Scene 1 denotes the opening scene and each cut increments the count by one. Shaded bands indicate $\pm$1 SD. App videos accrue scenes more rapidly, reaching ~4 distinct scenes by the end versus <2 for API videos, consistent with higher scene-cut frequency (App M = 3.10 cuts vs. API M = 0.76; d = 1.39, q < .001).}
\label{fig:temporal}
\end{figure}

\subsubsection{Audio and Speech}

Audio properties diverged across modalities. App videos were slightly louder (M = 0.096 $\pm$ 0.009 vs. 0.087 $\pm$ 0.013; d = 0.83, q < .001) but featured lower-pitched audio (M = 115.51 $\pm$ 30.04 Hz vs. 148.87 $\pm$ 40.78 Hz; d = $-$0.92, q < .001). Lower pitch in App videos may reflect the predominance of male narrators, different ambient sound profiles, or compositional choices in generated music.

Speech content also differed. API videos used first-person pronouns at nearly twice the rate of App videos (M = 0.078 $\pm$ 0.048 vs. 0.044 $\pm$ 0.039; d = $-$0.78, q < .001). API videos also contained more light-associated words (M = 0.16 $\pm$ 0.37 vs. 0.02 $\pm$ 0.14; d = $-$0.50, q = .03). Sentiment valence did not differ significantly between modalities (q = .63), with both producing mildly positive compound sentiment scores (App: M = 0.19 $\pm$ 0.34; API: M = 0.24 $\pm$ 0.37).

\subsubsection{Semantic Content}

Word frequency analysis revealed a coherent thematic vocabulary across the corpus (Figure 4). The most frequent terms clustered around emotional experience (``feel,'' n = 43; ``heavy,'' n = 25), temporal struggle (``day,'' n = 28; ``morning,'' n = 10), natural metaphors (``storm,'' n = 8; ``rain,'' n = 8; ``cloud,'' n = 8), and hope or release (``light,'' n = 13; ``breathe,'' n = 8).

\begin{figure}[htbp]
\centering
\includegraphics[width=\textwidth]{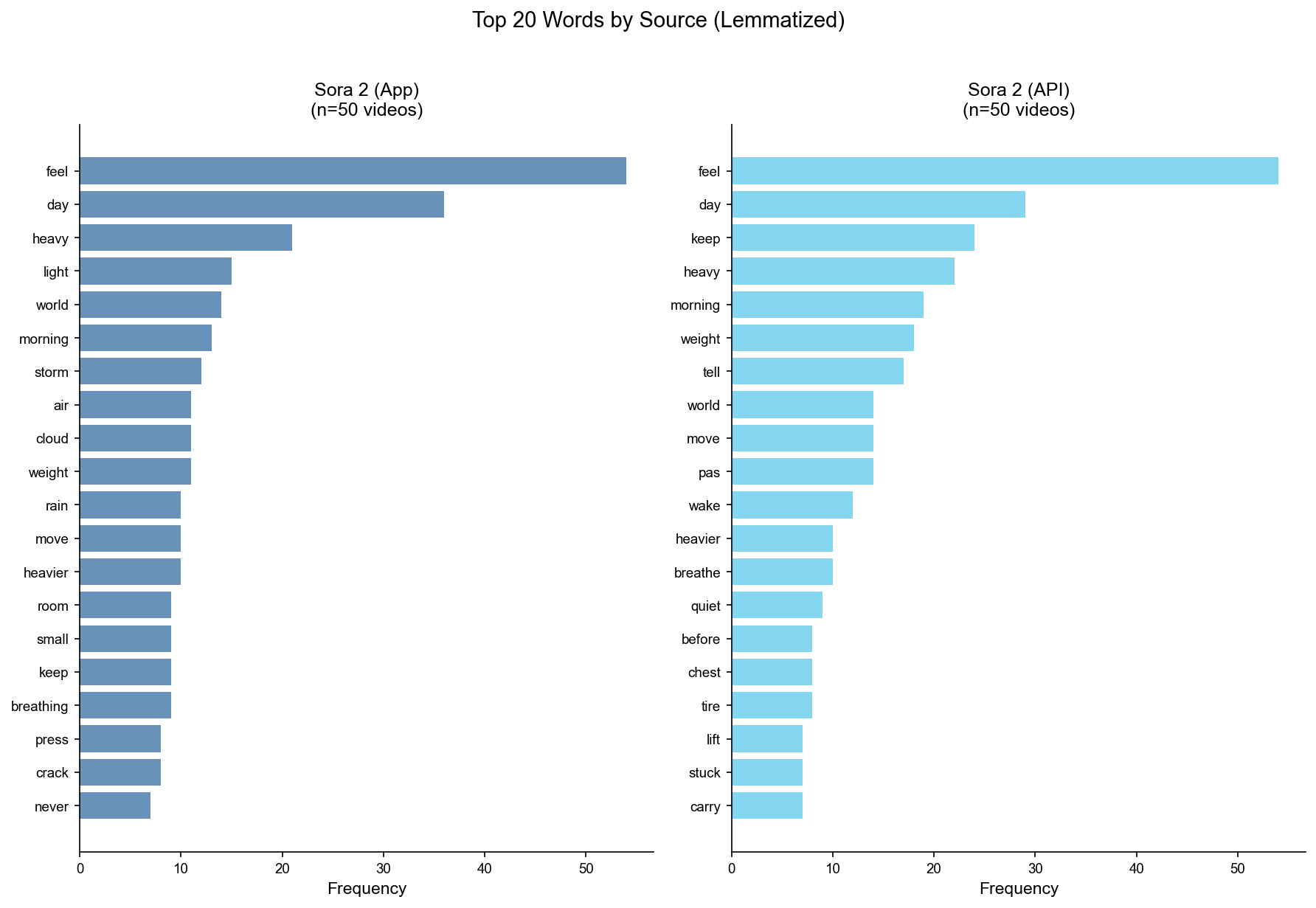}
\caption{Word frequency comparison. Twenty most frequent content words appearing in transcribed speech across App and API generated videos. Common terms cluster around themes of emotional experience (feel, heavy, weight), temporal struggle (day, morning), natural metaphors (storm, rain, cloud), and hope or release (light, breathe). This vocabulary reflects culturally prevalent metaphors for depression emphasizing weight, weather, darkness, and the possibility of relief.}
\label{fig:wordfreq}
\end{figure}

\subsection{Qualitative Findings}

\subsubsection{Inter-Rater Reliability}

Two coders independently coded all 100 videos. Reliability was strong for most dimensions: narrative arc ($\kappa$ = 0.69), environments ($\kappa$ = 0.91, IoU = 0.84), posture ($\kappa$ = 0.89), alone-in-frame ($\kappa$ = 0.94), and gender ($\kappa$ = 1.00). Object coding showed moderate overall agreement ($\kappa$ = 0.61) with variation by object type. Dimensions with insufficient reliability were excluded from analysis: facial expression ($\kappa$ = 0.49), race/ethnicity ($\kappa$ = 0.53), and apparent SES ($\kappa$ = 0.20). For objects, lower kappa values for ubiquitous items like beds ($\kappa$ = 0.44) and windows ($\kappa$ = 0.47) reflected asymmetric exhaustiveness between coders rather than substantive disagreement about object presence; the primary coder tagged these common items more exhaustively while the secondary coder focused on more distinctive elements. Full reliability results appear in Appendix E.

\subsubsection{Narrative and Visual Patterns}

The most striking qualitative finding concerned narrative structure. After collapsing to binary classification (recovery vs. no recovery, combining deterioration and no shift), a dramatic divergence emerged between modalities. App videos frequently depicted recovery narratives with 39 of 50 videos (78\%) containing a discernible shift toward hope or relief. API videos showed the opposite pattern: only 7 of 50 videos (14\%) depicted recovery arcs. This aligns with quantitative trajectory analysis showing App videos brightening over time while API videos remained flat.

\begin{figure}[htbp]
\centering
\includegraphics[width=\textwidth]{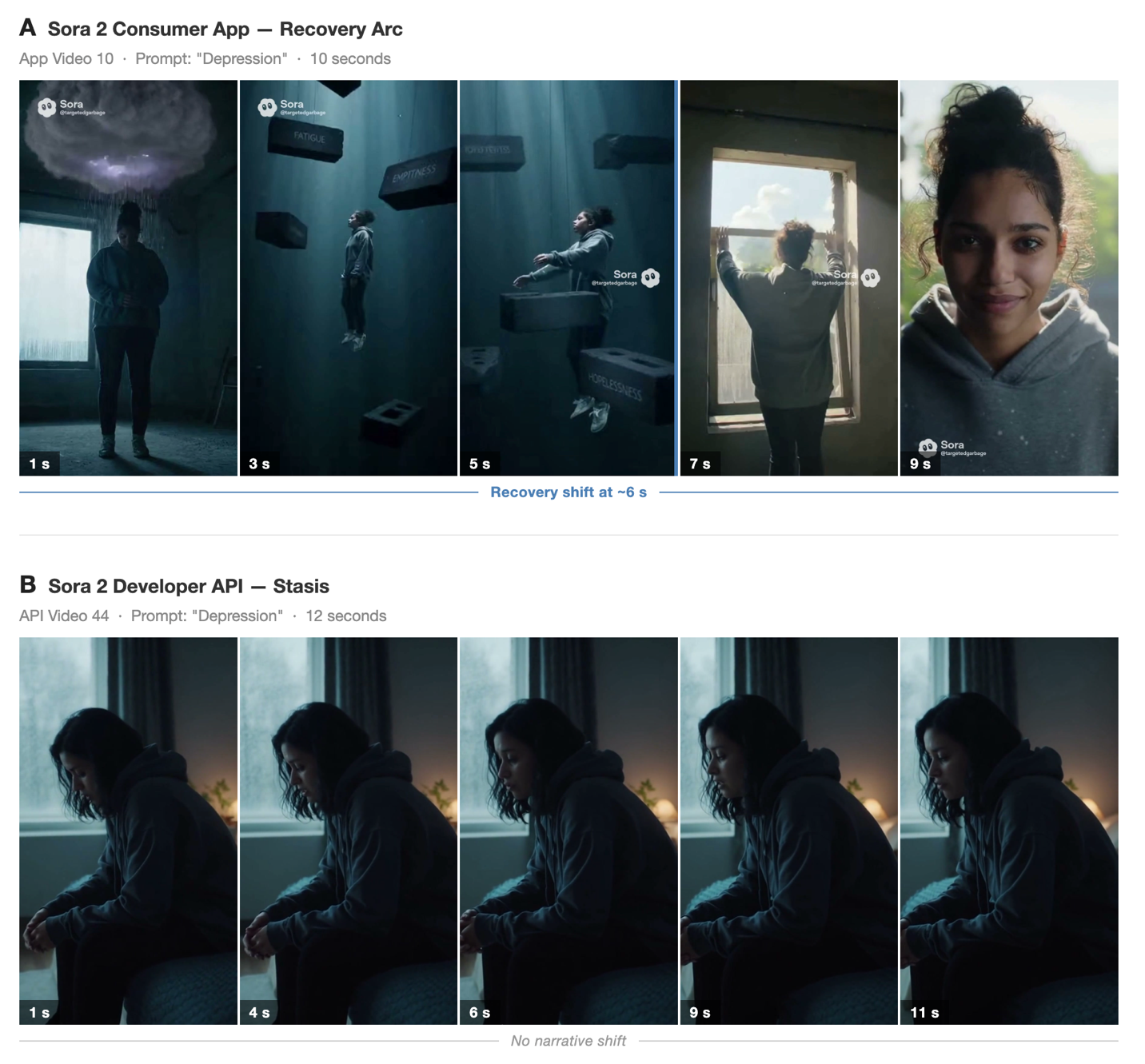}
\caption{Representative video stills from App and API outputs generated with the prompt ``Depression.'' (A) Consumer App video showing a typical recovery arc. The video progresses from a dark interior with a personal storm cloud (1 s), through floating debris labeled with terms like ``fatigue'' and ``hopelessness'' (3--5 s), to the figure turning toward a bright window (7 s) and ending outdoors in sunlight, smiling (9 s). The coded recovery shift occurs at approximately 6 seconds. (B) Developer API video showing typical stasis. The figure remains seated on a bed in a dim bedroom throughout the full 12-second duration, wearing a grey hoodie with downward gaze, clasped hands, and minimal change in posture, lighting, or environment. Both videos were generated with identical single-word prompts and no additional parameters.}
\label{fig:stills}
\end{figure}

Across both modalities, videos converged on a narrow visual vocabulary. Bedroom was the dominant setting, with App videos featuring somewhat more diverse environments including elevated outdoor spaces and urban settings. The most common objects were hoodies (n = 194), windows (148), beds (114), rain (83), and lamps (59).

Generated figures were predominantly young adults aged 20--30 (88\%) and nearly always alone in frame with downward gaze-- which in recovery videos often shifted upward. App videos showed greater physical mobility (66\% containing standing, walking, or floating) compared to API videos (12\%). Gender differed by modality: App outputs skewed male (68\%), API outputs skewed female (59\%).

\subsubsection{Recovery Transition Patterns}

To examine whether recovery patterns extended beyond coded objects and environments, we analyzed transcript words and quantitative visual features before versus after coded recovery timestamps (n = 46 videos with recovery arcs).

Analysis of environment appearance relative to recovery shifts (n = 46 videos) revealed systematic patterns. Environments appearing predominantly before recovery shifts included Bedroom (26 videos before vs. 3 after), Bathroom (7 vs. 0), and Urban Outdoor (12 vs. 4). Elevated outdoor settings (rooftops, hilltops, bridges) showed the opposite pattern (5 videos before vs. 8 after), appearing predominantly after recovery shifts. Objects followed parallel trajectories. Beds (32 vs. 1), rain (23 vs. 1), windows (32 vs. 8), and curtains (12 vs. 2) clustered in depressive phases, while sunlight and sunrise appeared predominantly after recovery shifts (3 vs. 17), as did birds (1 vs. 7). In terms of figure posture, gaze shifted from downward to upward in 68\% of recovery videos compared to 16\% of non-recovery videos.

Transcript vocabulary reinforced these patterns. Words appearing predominantly before recovery timestamps included ``heavy'' (n = 21), ``feels'' (n = 25), ``heavier'' (n = 10), ``room'' (n = 8), ``world'' (n = 14), ``rain'' (n = 7), and ``storm'' (n = 7). Words appearing predominantly after recovery timestamps included ``light'' (n = 12), ``small'' (n = 7), ``clouds'' (n = 5), ``breath'' (n = 5), ``cracks'' (n = 4), and ``moment'' (n = 4). Depression-phase vocabulary clustered around weight/pressure metaphors (heavy, heavier, pressing, chest) and water/confinement imagery (underwater, drowning, room), while recovery-phase vocabulary emphasized light, smallness, and atmospheric imagery.

Quantitative visual features showed corresponding patterns. Comparing mean values before versus after recovery timestamps: brightness increased by 27.4\% (d = 0.70, p < .001), saturation decreased by 13.5\% (d = $-$0.42, p < .001), and colorfulness increased by 10.8\% (d = 0.28, p = .002). Motion showed no significant change (p = .98).

\begin{figure}[htbp]
\centering
\includegraphics[width=0.9\textwidth]{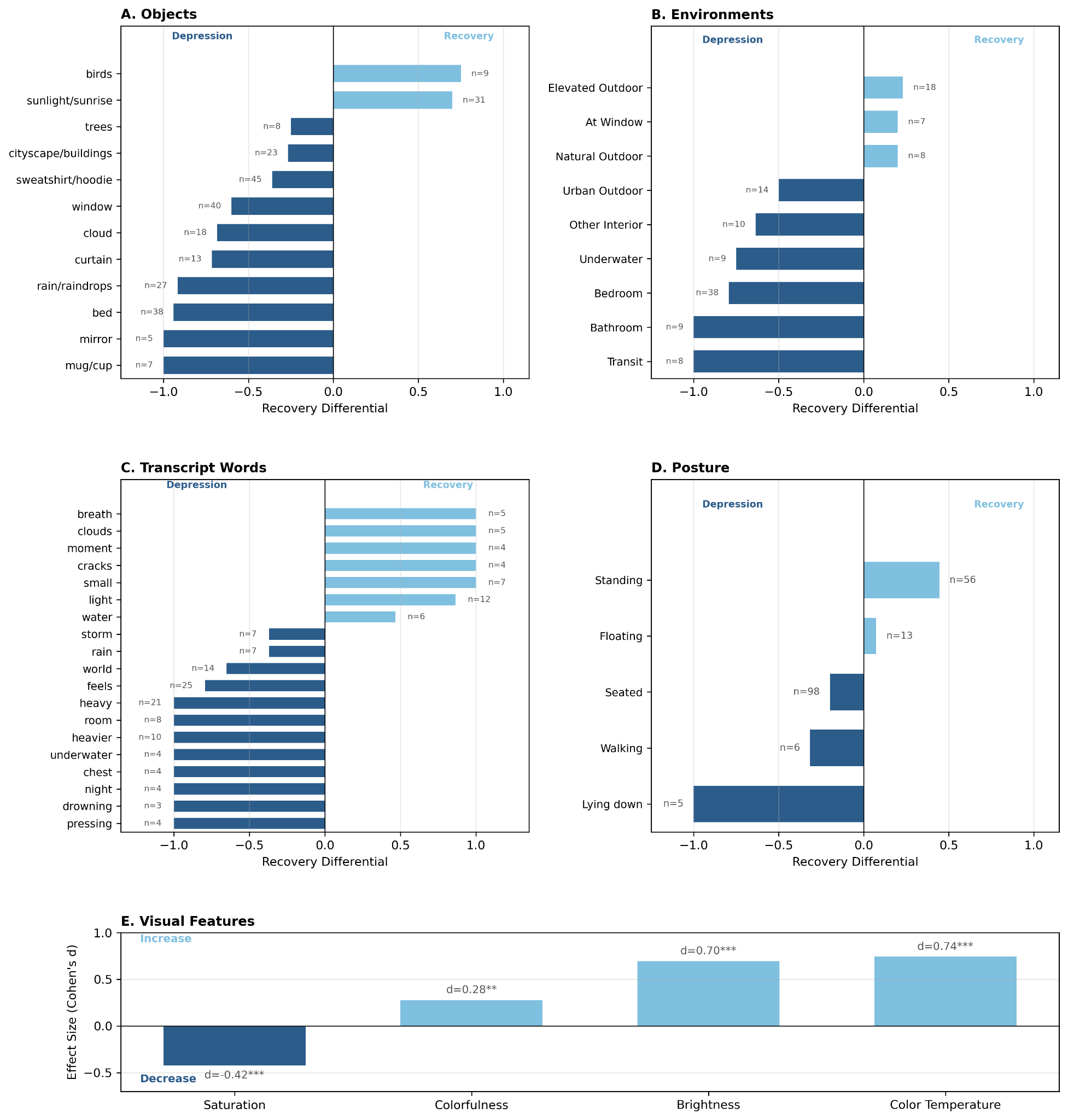}
\caption{\small Recovery-linked content shifts across objects, settings, language, posture, and low-level visual features. Panels A--D show recovery differentials comparing element presence (or rates) after vs. before the coded recovery timestamp in videos with recovery arcs (46 videos). Recovery differential was computed as (after$-$before)/(after+before); positive values (light blue) indicate elements appearing predominantly after the recovery shift, and negative values (dark blue) indicate elements appearing predominantly before the shift. Bar labels indicate the number of videos containing each coded element (A, B, D) or total word instances (C). (A) Objects and (B) Environments meeting inclusion thresholds (n $\geq$ 5) are shown; objects were additionally selected for thematic relevance. (C) Transcript words were chosen a priori by investigators for semantic relevance to depression and recovery themes and plotted using recovery differential based on before/after word rates. (D) Posture codes show recovery-linked changes in embodied state (e.g., lying down vs. standing/active positions). (E) Visual features summarize changes in saturation, colorfulness, brightness, and color temperature, expressed as effect size (Cohen’s d) for after-versus-before differences; positive values indicate increases and negative values indicate decreases (significance markers as shown).}
\label{fig:recovery}
\end{figure}

\section{Discussion}

Our findings indicate that depictions of depression generated by the Sora 2 App depend on more than the underlying model. How that model is embedded, constrained, and circulated in the product/app has a direct impact on the content, aesthetics, and narrative preferences present in outputs; in ways that diverge from base model behaviors. When prompted with the isolated diagnostic term ``Depression,'' the consumer-facing Sora app overwhelmingly produces recovery-oriented narratives, while the same model accessed through the API yields largely static, unresolved scenes. 78\% of app-generated videos contained a discernible shift toward hope or relief, compared with only 14\% of API outputs.

The contrast between product/app and API outputs indicates that product-layer mediation shapes how depression is represented. The Sora app positions video generation within a social feed, governed by community guidelines that prohibit content perceived as promoting depression or depicting self-harm \cite{ref8}. Whether this recovery shift bias stems from these explicit content safeguards, or emerges indirectly from platform dynamics that favor narrative progression and engagement cannot be determined from our analysis. Regardless of mechanism, our analysis suggests that the Sora 2 product layer introduces strong narrative constraints. Within this environment, outputs are disproportionately oriented toward recovery as a safe and legible endpoint. Clinically, this suggests that model outputs can be steered towards specific behaviors, and that clinical teams can play a role in improving AI video tools at the product level.

Our results are consistent with broader concerns that generative systems may favor culturally overrepresented narrative endpoints (e.g., happy endings, resolved arcs, completed transformations) when responding to ambiguous prompts rather than sustaining unresolved middle states. Writers working with language models have described this tendency toward narrative overfitting \cite{ref39}, noting in particular a bias toward positive narrative endings \cite{ref40}. In video generation, this bias manifests as a rapid exit from stasis into motion, light, elevation, and relief. This pattern may stem from training objectives, or it may encode the dominance of the ``restitution narrative'' in human-generated content: the culturally preferred illness storyline in which sickness is a temporary interruption resolved through recovery \cite{ref41}. This dynamic is not absent in the API’s outputs, but it is substantially amplified when the model operates within the Sora 2 App, which is effectively a social distribution pipeline optimized for engagement, shareability, and moderation.

Importantly, the visual and narrative grammars employed in these videos are not novel. The withdrawn figure in a dim bedroom, the rain-streaked window, and the gray hoodie all echo iconographies of melancholia traceable from Renaissance allegory and Enlightenment-era psychiatric imaging through to modern media depictions, consistent with patterns observed in text-to-image models \cite{ref23}. The visual grammar also aligns with existing research on depression metaphors that identify darkness, descent, weight, and containment as near-universal conceptual frameworks \cite{ref42,ref43,ref44}. Our findings seem to confirm these patterns' persistence into AI video models. The dominance of solitary, seated figures, the clustering of beds, rain, and windows in depressive phases, and the systematic reversal toward light, birds, and elevation during recovery all make use of established symbols and motifs from media. Our findings suggest that Sora 2 does not invent new ways of depicting depression; instead, it compresses and recombines historical visual conventions and routes them through contemporary platform logics that privilege certain endpoints and user behaviors over others. Clinicians should be aware that AI-generated content about depression reflects training data and platform design rather than clinical expertise, and may need to help patients contextualize the simplified narratives and homogeneous imagery they encounter. For researchers and policymakers, these findings reiterate a methodological imperative: analyses of generative AI must attend to both base models, and the media forms and distribution systems through which AI model outputs are encountered by public userbases.

This study has several limitations. The analysis is restricted to a single platform (Sora 2) and a single prompt term, limiting generalizability to other systems or prompt formulations. The disparity in default clip length between access points (10 seconds for App, 12 seconds for API) introduces a potential confound. Where possible, we mitigated this by analyzing per-second rates and slopes rather than raw totals, but the difference cannot be fully disentangled from product-layer effects. Coders were not blinded to access modality, as the generation interfaces differed visibly, which may have introduced expectation effects in coding. Finally, the sample of 50 videos per modality was chosen to establish preliminary baseline observations rather than to power specific inferential comparisons; effect sizes and confidence intervals should be interpreted accordingly.

\section{Conclusion}

This study provides the first systematic analysis of how the Sora 2 generative video model depicts depression, revealing that its outputs are shaped as much by product-layer mediation as by underlying model associations. The consumer App's pronounced recovery bias (with 78\% of videos progressing toward hope compared to only 14\% of API outputs) demonstrates that platform constraints actively reshape which mental health narratives reach users. Across both access points, the visual and linguistic grammar proved remarkably consistent with centuries-old iconographic traditions: the solitary figure, the dim interior, the downward gaze, the metaphorics of weight and weather. These findings suggest that Sora 2 predominantly recirculates familiar cultural conventions for depicting depression rather than departing sharply from them. We expect that this pattern may well extend to other generative video models built on similar architectures and training data. As AI-generated content becomes increasingly prevalent in the informational environments of people experiencing psychiatric distress, clinicians should understand that these outputs reflect training data and platform constraints rather than clinical knowledge, and that they may shape patient expectations about recovery timelines and trajectories. Researchers and policymakers, in turn, should attend to the distinction between model behavior and product behavior when evaluating these systems.

\section*{Acknowledgments}

Figure 1 was generated using Google Paper Banana, a research figure generative AI tool; all design decisions and content were specified by the authors. Claude Opus 4.6 (Anthropic) was used for proofreading and copy-editing during manuscript preparation. All scientific content, analysis, interpretation, and final text were authored and reviewed by the authors.

\section*{Funding}

This research received no specific grant from any funding agency in the public, commercial, or not-for-profit sectors.

\section*{Conflicts of Interest}

None declared.

\section*{Data Availability}

The analysis code and video dataset analyzed in this study are available from the corresponding author upon reasonable request.

\section*{Authors' Contributions}

MF: Conceptualization, Methodology, Software Development, Formal Analysis, Investigation, Data Curation, Writing - Original Draft, Writing - Review \& Editing, Visualization, Project Administration. GS: Conceptualization, Methodology, Writing - Review \& Editing. JH: Investigation, Writing - Review \& Editing. ZZ: Investigation, Data Curation. JT: Supervision, Writing - Review \& Editing.

\section*{Abbreviations}

\begin{description}
\item[API] application programming interface
\item[FDR] false discovery rate
\item[GLCM] gray-level co-occurrence matrix
\item[IoU] intersection over union
\item[SES] socioeconomic status
\item[SFV] short-form video
\item[VADER] Valence Aware Dictionary and Sentiment Reasoner
\end{description}

\bibliographystyle{unsrtnat}
\bibliography{references}

\begin{appendices}

\section{Extended Qualitative Methods}

\noindent\textbf{Coding procedure.} Each video was reviewed in three passes. The first pass was an orientation viewing without pausing to get an overall sense of narrative arc, mood, and content. The second pass involved detailed coding with pausing and frame-by-frame scrubbing as needed to tag all elements with accurate timestamps. The third pass was a verification review to confirm accuracy, fill gaps, and add notes about uncertainties. Coders were instructed to prioritize precision over interpretation (tagging what was directly observable rather than inferred) and to err toward over-tagging when uncertain.

\noindent\textbf{Coding dimensions.} We coded six dimensions for each video, each selected based on established findings in the depression representation literature.

Narrative arc captured whether the video presented a discernible shift in emotional trajectory and, if so, its direction: recovery (movement toward more positive, hopeful, or lighter presentation) or deterioration (movement toward more negative, darker, or heavier presentation). Shifts could be signaled through changes in lighting, color palette, music, figure posture, voiceover language, or environmental elements. Not all videos contained shifts; some presented a consistent mood throughout. This dimension addresses whether Sora 2 tends to depict depression as a static state, a worsening condition, or something that improves. Visual depictions of depression consistently employ trajectory metaphors, either escape from confinement or progressive entrapment, making directional shifts a key marker of how the condition is visually conceptualized \cite{ref45}.

Environments were coded with timestamps to capture all physical settings or spaces depicted in each video. We were interested in what kinds of environments Sora 2 associates with depression; whether predominantly indoor, outdoor, domestic, institutional, natural, urban, or abstract. Coders described environments as specifically as the video allowed (e.g., ``small bedroom with drawn curtains,'' ``forest path with fog,'' ``geometric void with floating shapes''), using general categories only when more specific description was not possible. Confinement and containment are among the most frequently documented metaphors for depression, often rendered visually through enclosed or restrictive spaces \cite{ref45}.

Objects were inventoried with timestamps to capture what visual symbols and props Sora 2 associates with depression. Categories of interest included furniture and fixtures, weather and atmospheric elements, nature, personal items, domestic objects, and symbolically notable items such as empty chairs, wilting plants, or broken glass. Visual symbols of depression draw on embodied metaphor systems linking the condition to darkness, descent, weight, and fragility \cite{ref45}.

Figure demographics captured information about human figures depicted in the videos, including apparent gender presentation, race/ethnicity, age (coded by decade), and apparent socioeconomic status (inferred from clothing, grooming, and setting context). This dimension may help reveal whether the model reproduces stereotypical representations of who experiences depression, though as discussed in Appendix E, reliability for some demographic variables proved insufficient for substantive analysis. Figures were tagged even when only partially visible (hands, silhouette, back of head), with visibility limitations noted. Voiceover narration without a visible speaker was treated as a figure, with demographic fields marked ``Unclear'' when not identifiable from voice alone. Stock imagery of depression disproportionately features young women and carries a more positive emotional tone when depicting men, patterns that may or may not resurface in AI generated videos \cite{ref46}.

Figure states captured how figures appeared and behaved over time, including posture (lying down, seated, standing, walking), facial expression, and whether the figure appeared alone in the frame. Because these attributes can change during a figure's appearance (e.g., a figure who begins lying in bed looking sad and ends standing by a window looking neutral) this dimension allowed multiple timestamped entries per figure to capture transitions. Facial expressions were coded using standardized categories: sad, happy/smiling, angry, anxious/worried, neutral/flat, tearful, pensive, tired/exhausted, distressed, unclear, or not visible. Slumped posture and social isolation are characteristic of both clinical presentation and media clich\'{e}; coding these features allows assessment of whether generated imagery reproduces narrow visual conventions \cite{ref47}.

Overall notes captured general observations, AI artifacts (visual glitches, morphing, distortions), and uncertainties requiring discussion. The full codebook with detailed definitions and edge case handling appears in Appendix B.

\noindent\textbf{Reliability.} Two authors (MF and ZZ) independently coded the full dataset using the structured codebook. One author's coding (MF) was designated as the primary dataset for analysis; the second author's coding (ZZ) served as inter-rater reliability verification.

Raw codes were harmonized prior to reliability analysis to reconcile minor vocabulary differences between coders. Environment descriptions were mapped to 11 standardized categories (Bedroom, Bathroom, Kitchen/Dining, Other Interior, At Window, Transit, Urban Outdoor, Natural Outdoor, Elevated Outdoor, Underwater, Abstract). Object descriptions were harmonized to standardized forms (e.g., ``sweatshirt,'' ``hoodie,'' ``sweater'' → ``sweatshirt/hoodie''); reliability was calculated at this harmonized object level. Facial expressions were collapsed into three affect categories: Positive (happy, peaceful, hopeful), Subdued/Low (tired, neutral, pensive), and Negative (sad, tearful, worried, distressed, angry). Posture was harmonized to six categories (Seated, Standing, Lying down, Walking, Floating/Swimming, Kneeling/Crouching). Minor variations in demographic coding (e.g., ``Medium'' vs. ``Middle'' for SES) were reconciled during harmonization. The full harmonization dictionaries appear in Appendix C.

Cohen's kappa ($\kappa$) was calculated for presence/absence of categorical elements within each video, including narrative direction, environment categories, harmonized objects, figure demographics, and figure states. For dimensions involving temporal segments (environments, objects, figure states), we also calculated intersection-over-union (IoU) at one-second resolution to assess agreement on when elements appeared. Coders frequently noted gaze direction in free-text annotations; agreement on gaze was assessed for videos where both coders documented directional looking behavior. Given the scale of the dataset, item-by-item consensus resolution was not feasible; reliability metrics verified that the coding scheme could be applied consistently across coders.

\section{Qualitative Tagging Guide}

\subsection*{Study Goal}

This study examines how OpenAI's Sora 2 video generation model visually interprets and represents the concept of depression. We generated 100 videos using the prompt ``Depression'' and are now systematically analyzing the visual, narrative, and demographic patterns that emerge.

Your role as a qualitative coder is to capture timestamped observations about what appears in each video. This qualitative data will be paired with quantitative computer vision analysis to establish a baseline behavior profile for how Sora 2 represents depression. The ultimate goal is to understand whether the model reproduces, reinforces, or challenges existing stereotypes and media conventions around mental health.

\subsection*{Overall Principles}

\noindent\textbf{Precision over interpretation: }Tag what you directly observe, not what you infer or interpret. If you see a person lying in bed, tag ``person lying in bed''---do not tag ``person who is exhausted'' unless exhaustion is explicitly indicated. When uncertain whether something qualifies, tag it and note your uncertainty in the Notes column.

\noindent\textbf{Timestamps: }All timestamps are in whole seconds from video start (0 = first frame). Round to the nearest second.

\noindent\textbf{Err toward over-tagging: }It is easier to filter out unnecessary data during analysis than to re-watch all 100 videos. When in doubt about whether something is notable, tag it.

\noindent\textbf{Inter-rater process: }Each video will be reviewed by at least two coders independently. Given the scale of the dataset, item-by-item consensus resolution is not feasible; and a primary coder's data will be used for analysis, with inter-rater reliability reported separately. Insufficiently reliable fields will be excluded from substantive analysis.

\subsection*{Watching Process}

\noindent\textbf{First pass (orientation): }Watch the full video without pausing to get an overall sense of the narrative arc, mood, and content. Take mental note of any shifts in tone or presentation.

\noindent\textbf{Second pass (detailed tagging): }Pause and scrub as needed to tag all elements across all tabs (Narrative, Environments, Objects, Figures) with accurate timestamps.

\noindent\textbf{Third pass (verification): }Review your tags against the video to confirm accuracy, fill in any gaps, and add notes about uncertainties or edge cases.

\bigskip
\noindent\rule{\textwidth}{0.4pt}

\subsection*{Tab 1: Narrative}

This tab captures whether each video presents a narrative trajectory. Specifically, whether the emotional or thematic direction of the video changes over its duration. We are interested in whether Sora 2 tends to show depression as a static state, a worsening condition, or something that improves.

\begin{table}[H]
\centering
\small
\caption{Column descriptions for Narrative tab.}
\label{tab:coding_0}
\begin{tabular}{l p{0.65\textwidth}}
\toprule
Column & Description \\
\midrule
Video ID & Filename of the video (e.g., ``sora2\_depression\_001.mp4'') \\
Shift Present? & Yes / No: Does the video contain a discernible narrative shift? \\
Shift Second & Timestamp (in seconds) where the shift begins. Leave blank if Shift Present = No. \\
Direction & ``Recovery'' / ``Deterioration'': The direction of the shift. Leave blank if Shift Present = No. \\
Notes & Any clarifying notes or description that are worth calling out \\
\bottomrule
\end{tabular}
\end{table}

\paragraph{Key Concept: Narrative Shift.}
A ``shift'' is a discernible change in the video's emotional or thematic direction. This could be signaled through changes in lighting, color palette, music, figure posture, voiceover language, or environmental elements. Not all videos will contain a shift. Some may present a consistent mood throughout.

\paragraph{Direction Definitions.}~

\begin{itemize}
\item \textbf{Recovery:} A shift toward more positive, hopeful, or lighter presentation. Examples include: lighting warming or brightening; a figure standing up, going outside, or engaging with others; music becoming more uplifting; voiceover expressing hope or improvement.
\item \textbf{Deterioration:} A shift toward more negative, darker, or heavier presentation. Examples include: lighting dimming or shifting to cooler tones; a figure withdrawing, lying down, or isolating; music becoming more somber; voiceover expressing despair or worsening.
\end{itemize}

\paragraph{Indicators to watch for.}
When determining whether a shift has occurred and its direction, pay attention to the following types of signals. If any of these are particularly notable, describe them in the Notes field:
\begin{itemize}
\item Voiceover language: transitional phrases like ``and yet,'' ``but sometimes,'' ``until one day''
\item Visual shifts: lighting changes, color warming/cooling, figure posture changes
\item Audio shifts: music tone changes, volume shifts
\end{itemize}

\noindent\rule{\textwidth}{0.4pt}

\subsection*{Tab 2: Environments}

This tab captures the physical settings or spaces depicted in each video. We want to understand what kinds of environments Sora 2 associates with depression. Are they predominantly indoor, outdoor, domestic, institutional, natural, urban, or abstract?

\begin{table}[H]
\centering
\small
\caption{Column descriptions for Environments tab.}
\label{tab:coding_1}
\begin{tabular}{l p{0.65\textwidth}}
\toprule
Column & Description \\
\midrule
Video ID & Filename of the video (should be the same across all tabs for the same video) \\
Start Second & Timestamp when this environment first appears \\
End Second & Timestamp when this environment ends (or video ends) \\
Environment & Description of environment (see guidance below) \\
Notes & Any additional salient details about the environment/setting \\
\bottomrule
\end{tabular}
\end{table}

\paragraph{Environment Description Guidance.}
Describe the environment as specifically as the video allows. Be concrete and descriptive rather than interpretive.

\noindent\textbf{Good examples: }``subway car at night,'' ``hospital waiting room,'' ``forest path with fog,'' ``open field at dusk,'' ``small bedroom with drawn curtains,'' ``geometric void with floating shapes,'' ``bathroom with running shower.''

When a more specific description isn't possible, use these general categories:
\begin{itemize}
\item \textbf{Indoor:} Generic interior space where specific room type is unclear
\item \textbf{Outdoor:} Generic natural landscape (field, forest, sky, water) without more specific features
\item \textbf{Urban:} Generic city or built environment (streets, buildings, infrastructure)
\item \textbf{Abstract:} Non-representational space (geometric patterns, voids, surreal landscapes)
\end{itemize}

If the setting is ambiguous or transitioning, describe what you observe and note uncertainty in the Notes column.

\noindent\rule{\textwidth}{0.4pt}

\subsection*{Tab 3: Objects}

This tab captures non-human objects that appear in the videos. We are interested in what visual symbols and props Sora 2 associates with depression: medication bottles, rain, empty chairs, phones, etc.

\begin{table}[H]
\centering
\small
\caption{Column descriptions for Objects tab.}
\label{tab:coding_2}
\begin{tabular}{l p{0.65\textwidth}}
\toprule
Column & Description \\
\midrule
Video ID & Filename of video \\
Start Second & Timestamp when the object first appears in frame \\
End Second & Timestamp when the object leaves the frame (or video ends) \\
Object & Name of the object \\
Notes & Adjectives or other relevant qualifiers (e.g., color, state, condition) \\
\bottomrule
\end{tabular}
\end{table}

\paragraph{Tagging Guidance.}
Tag all reasonably identifiable objects that appear in the video, including clothing, furniture, fixtures, and environmental elements. Use the Notes column to capture relevant qualifiers: state (lamp: ``On'' / ``Off''), color (sweatshirt: ``Grey''), condition (plant: ``Living'' / ``Wilting''), or other descriptors. If an object appears, disappears, and reappears, tag each appearance as a separate row.

\paragraph{Object Reference List.}
This list is not exhaustive; describe freely as needed. It is meant to calibrate you to the types of objects we consider notable:
\begin{itemize}
\item \textbf{Furniture \& fixtures:} bed, chair, couch, table, desk, window, mirror, door, lamp, stairs, bathtub, shower
\item \textbf{Weather \& atmospheric:} rain, raindrops on glass, clouds, storm, fog, mist, snow, sunshine, god-rays, lightning
\item \textbf{Nature:} trees, forest, ocean, water, river, lake, flowers, plants, grass, field, mountain, sky, moon, sun, stars
\item \textbf{Personal items:} phone, medication/pill bottle, book, journal, photographs, cup/mug, food, cigarette, headphones
\item \textbf{Domestic objects:} blanket, pillow, curtains, clock, television, computer, dishes
\item \textbf{Symbolic/notable:} empty chair, wilting plant, broken glass, candle, shadow, reflection
\end{itemize}

\noindent\rule{\textwidth}{0.4pt}

\subsection*{Tab 4: Figures}

This tab captures information about human figures (and voices) depicted in the videos. We are interested in the demographics of who Sora 2 depicts when visualizing depression, as well as how these figures are portrayed (posture, expression, isolation).

\begin{table}[H]
\centering
\small
\caption{Column descriptions for Figures tab.}
\label{tab:coding_3}
\begin{tabular}{l p{0.65\textwidth}}
\toprule
Column & Description \\
\midrule
Video ID & Filename of video \\
Figure ID & Figure 1, Figure 2, etc. (in order of first appearance) \\
Start Second & Timestamp when figure first appears \\
End Second & Timestamp when figure leaves frame \\
Gender & Observed gender presentation: Man / Woman / Nonbinary / Unclear \\
Race/Ethnicity & Observed race/ethnicity, or ``Unclear'' if not identifiable \\
Age & Decade estimate (teens, 20-30, 30-40, 40-50, 50-60, 60+, unclear) \\
Apparent SES & Inferred socioeconomic status based on appearance: Low / Middle / High / Unclear (see guidance) \\
Notes & Visibility limitations, posture, or other relevant observations \\
\bottomrule
\end{tabular}
\end{table}

\paragraph{Tagging Guidance.}~
\begin{itemize}
\item \textbf{Partial visibility:} Tag figures even if only partially visible (hands, silhouette, back of head). Note the limitation in the Notes column (e.g., ``only hands visible,'' ``silhouette only'').
\item \textbf{Continuity across cuts:} If the same figure is clearly continuous across a cut, use the same Figure ID. If it's uncertain whether it's the same person, assign a new ID and note the uncertainty.
\item \textbf{Voiceovers:} Treat voiceover narration as a figure, using ``Voice 1,'' ``Voice 2'' as the Figure ID. Demographic fields may be marked ``Unclear'' if not identifiable from voice.
\item \textbf{AI distortions:} If AI artifacts distort a figure, still attempt to tag what you can observe and note the distortion.
\end{itemize}

\paragraph{Apparent SES Guidance.}
This is inherently a rough inference. Base your judgment on observable cues such as:
\begin{itemize}
\item Clothing quality and style (worn/unkempt vs. well-maintained vs. expensive-looking)
\item Grooming and personal presentation
\item Setting context (sparse room vs. well-furnished home vs. affluent environment)
\item Mark ``Unclear'' if there are insufficient cues to make even a rough judgment.
\end{itemize}

\noindent\rule{\textwidth}{0.4pt}

\subsection*{Tab 5: Figure States}

This tab captures how figures appear and behave over time: their posture, facial expression, and social context within the frame. Because these attributes can change during a figure's appearance (e.g., a figure who begins lying in bed looking sad and ends standing by a window looking neutral), this tab allows multiple entries per figure to capture those transitions.

\begin{table}[H]
\centering
\small
\caption{Column descriptions for Figure States tab.}
\label{tab:coding_4}
\begin{tabular}{l p{0.65\textwidth}}
\toprule
Column & Description \\
\midrule
Video ID & Filename of the video \\
Figure ID & Links to the figure in Tab 4 (e.g., ``Figure 1''). Must match an existing Figure ID. \\
Start Second & Timestamp when this state begins \\
End Second & Timestamp when this state ends or figure leaves frame \\
Posture & Lying down / Seated / Standing / Walking / Unclear \\
Facial Expression & See options below \\
Alone in Frame & Yes / No: Is this figure the only person visible during this state? \\
Notes & Qualifiers or context (e.g., ``head in hands,'' ``looking out window,'' ``back to camera,'' ``in bed'') \\
\bottomrule
\end{tabular}
\end{table}

\paragraph{Tagging Guidance.}
Create a new row whenever a figure's posture, expression, or social context (alone/not alone) meaningfully changes. Not every micro-movement requires a new entry. Use judgment. If a figure shifts from sad to neutral over a few seconds, that's worth capturing. If they blink or briefly glance away, it's not.

\paragraph{Facial Expression Guidance.}
Tag the predominant expression during the figure's appearance. Options:
\begin{itemize}
\item \textbf{Sad:} Downturned mouth, furrowed brow, drooping features
\item \textbf{Happy/Smiling:} Upturned mouth, raised cheeks, bright eyes
\item \textbf{Hopeful/Peaceful:} Relaxed face, soft eyes
\item \textbf{Angry:} Glaring eyes, clenched jaw, compressed lips, hard downward brow (intensity, aggression)
\item \textbf{Anxious/Worried:} Raised inner eyebrows, wide or darting eyes, lip biting, tense but uncertain (unease, apprehension)
\item \textbf{Neutral/Flat:} Blank or expressionless affect
\item \textbf{Tearful:} Visible tears or crying
\item \textbf{Pensive:} Thoughtful, contemplative, gazing distantly
\item \textbf{Tired/Exhausted:} Heavy eyelids, slack features, weary appearance
\item \textbf{Distressed:} Visible anguish, tension, or agitation
\item \textbf{Unclear:} Face visible but expression hard to categorize
\item \textbf{N/A:} Face not visible (back of head, silhouette, hands only)
\end{itemize}

\noindent\rule{\textwidth}{0.4pt}

\subsection*{Tab 6: Notes}

This tab captures overall observations, uncertainties, edge cases, issues, or random notes that do not fit elsewhere in the tagging sheet. Every video should have exactly one row in this tab.

\begin{table}[H]
\centering
\small
\caption{Column descriptions for Notes tab.}
\label{tab:coding_5}
\begin{tabular}{l p{0.65\textwidth}}
\toprule
Column & Description \\
\midrule
Video ID & Filename of video \\
Date Coded & Date you coded this video \\
Notes & Free-text field for overall observations \\
\bottomrule
\end{tabular}
\end{table}

\noindent\rule{\textwidth}{0.4pt}

\subsection*{Handling Edge Cases}

\noindent\textbf{Abstract or non-representational content: }Tag what you can. Environments can be ``abstract.'' If figures are distorted or surreal, attempt demographic coding with ``Unclear'' where needed.

\noindent\textbf{AI artifacts and glitches: }If visual glitches, morphing, or distortions make identification difficult, note this in Tab 6 but still attempt to tag what you observe.

\noindent\textbf{No figures present: }If a video contains no human figures or voiceovers, leave the Figures tab empty for that video. This is valid data.

\noindent\textbf{Rapid scene changes: }If environments or objects change very rapidly (e.g., montage sequences), tag the most salient or longest-duration elements. Note in Tab 5 that the video contained rapid cuts.

\noindent\textbf{Unclear or borderline cases: }When genuinely uncertain, err on the side of tagging and note your uncertainty. Given the scale of the dataset, item-by-item consensus resolution is not feasible; reliability metrics will be used to verify coding consistency, and a primary coder's judgments will be used for analysis.

\section{Harmonization Dictionaries}

To reconcile minor vocabulary differences between coders and enable systematic analysis, raw codes were mapped to standardized categories. All mappings used case-insensitive substring matching unless otherwise noted.

\subsection{Environment Harmonization}

Notes: ``underwater bedroom'' mapped to Underwater. Close-up shots carried forward the previous environment.

\begin{table}[H]
\centering
\small
\caption{Environment Harmonization}
\label{tab:harmonize_0}
\begin{tabular}{l p{0.6\textwidth}}
\toprule
Standardized Category & Raw Code Patterns \\
\midrule
Bedroom & ``bedroom'', ``in bed'' \\
Bathroom & ``bathroom'', ``bathtub'', ``in bathtub'' \\
Kitchen/Dining & ``kitchen'', ``dining'', ``cafe'' \\
Other Interior & ``desk'', ``living room'', ``hallway'', ``basement'', ``empty room'', ``elevator'' \\
At Window & ``window'', ``at window'', ``next to window'', ``in front of window'' \\
Transit & ``bus'', ``subway'', ``train'', ``car'', ``transit'' \\
Urban Outdoor & ``street'', ``city'', ``urban'', ``underpass'', ``parking'' \\
Natural Outdoor & ``park'', ``field'', ``meadow'', ``river'', ``lake'', ``lagoon'', ``pond'', ``wheat'', ``flower'', ``garden'', ``ice flat'', ``outside'' \\
Elevated Outdoor & ``roof'', ``rooftop'', ``hilltop'', ``mountain'', ``balcony'', ``top of'', ``bridge'' \\
Underwater & ``underwater'', ``under water'', ``submerged'', ``in water'' \\
Abstract & ``space'', ``cloud'', ``abstract'', ``galaxy'', ``sky'', ``dome'' \\
\bottomrule
\end{tabular}
\end{table}

\subsection{Object Harmonization}

Note: Objects not matching any pattern were retained as-is.

\begin{table}[H]
\centering
\small
\caption{Object Harmonization}
\label{tab:harmonize_1}
\begin{tabular}{l p{0.6\textwidth}}
\toprule
Standardized Object & Raw Code Patterns \\
\midrule
sweatshirt/hoodie & ``sweatshirt'', ``hoodie'', ``hoddie'', ``sweater'', ``long-sleeve'', ``long sleeve'', ``t-shirt'', ``shirt'' \\
jacket/coat & ``jacket'', ``coat'', ``rainjacket'', ``raincoat'', ``blazer'' \\
bed & ``bed'' (excluding ``bedroom'') \\
couch/sofa & ``couch'', ``sofa'' \\
end table & ``end table'', ``nightstand'' \\
table/desk & ``table'', ``desk'' \\
rain & ``rain'', ``heavy rain'' \\
raindrops & ``raindrop'', ``rain on'' \\
cloud & ``cloud'' \\
sunrise/sunset & ``sunrise'', ``sunset'' \\
sunlight/god rays & ``god ray'', ``god-ray'', ``sunlight'' \\
standing water/puddles & ``standing water'', ``water on'', ``puddle'' \\
bathtub & ``bathtub'' \\
bubbles & ``bubble'' \\
plant/flowers & ``plant'', ``flower'', ``sunflower'' \\
trees & ``Tree'', ``trees'' \\
birds & ``bird'' \\
phone & ``phone'' \\
mug/cup & ``mug'', ``cup'' \\
lamp & ``lamp'' \\
window & ``window'', ``windows'' \\
curtain & ``Curtain'', ``curtains'' \\
tear & ``tear'' \\
mirror & ``mirror'' \\
hands & ``hands'', ``hand'' \\
cityscape/buildings & ``cityscape'', ``building'', ``skyline'' \\
cars & ``car'', ``cars'' \\
picture/poster & ``picture'', ``poster'', ``photo'', ``polaroid'' \\
bench & ``bench'' \\
\bottomrule
\end{tabular}
\end{table}

\subsection{Figure State Harmonization}

\begin{table}[H]
\centering
\small
\caption{Posture Harmonization}
\label{tab:harmonize_2}
\begin{tabular}{l p{0.6\textwidth}}
\toprule
Standardized Category & Raw Code Patterns \\
\midrule
Seated & ``seated'', ``sitting'', ``seted'' \\
Standing & ``standing'', ``standng'', ``stading'' \\
Lying down & ``lying'' \\
Walking & ``walking'' \\
Floating/Swimming & ``floating'', ``swimming'' \\
Kneeling/Crouching & ``kneeling'', ``crouching'', ``squatting'', ``crawling'' \\
\bottomrule
\end{tabular}
\end{table}

\begin{table}[H]
\centering
\small
\caption{Facial Expression Harmonization}
\label{tab:harmonize_3}
\begin{tabular}{l p{0.6\textwidth}}
\toprule
Affect Category & Raw Codes \\
\midrule
Positive & ``happy'', ``hopeful'', ``peaceful'', ``surprised'' \\
Subdued/Low & ``tired'', ``neutral'', ``pensive'', ``flat'', ``focused'', ``sleeping'' \\
Negative & ``sad'', ``tearful'', ``crying'', ``worried'', ``distressed'', ``anxious'', ``angry'', ``desperate'', ``dejected'' \\
\bottomrule
\end{tabular}
\end{table}

\begin{table}[H]
\centering
\small
\caption{Alone-in-Frame Harmonization}
\label{tab:harmonize_4}
\begin{tabular}{ll}
\toprule
Standardized & Raw Codes \\
\midrule
Yes & ``yes'' \\
No & ``no'' \\
\bottomrule
\end{tabular}
\end{table}

\subsection{Figure Demographic Harmonization}

\begin{table}[H]
\centering
\small
\caption{Apparent SES Harmonization}
\label{tab:harmonize_5}
\begin{tabular}{ll}
\toprule
Standardized & Raw Codes \\
\midrule
High & ``high'' \\
Medium & ``medium'', ``middle'' \\
Low & ``low'' \\
\bottomrule
\end{tabular}
\end{table}

\begin{table}[H]
\centering
\small
\caption{Race/Ethnicity Harmonization}
\label{tab:harmonize_6}
\begin{tabular}{ll}
\toprule
Standardized & Raw Codes \\
\midrule
Black & ``black'' \\
White & ``white'' \\
Latino/Hispanic & ``hispanic'', ``latino'', ``latina'' \\
Asian & ``asian'' \\
Arab/Middle Eastern & ``arab'', ``middle eastern'' \\
Indian & ``indian'' \\
\bottomrule
\end{tabular}
\end{table}

\begin{table}[H]
\centering
\small
\caption{Gender Harmonization}
\label{tab:harmonize_7}
\begin{tabular}{ll}
\toprule
Standardized & Raw Codes \\
\midrule
Male & ``male'' \\
Female & ``female'' \\
\bottomrule
\end{tabular}
\end{table}

\begin{table}[H]
\centering
\small
\caption{Age Harmonization}
\label{tab:harmonize_8}
\begin{tabular}{ll}
\toprule
Standardized & Raw Codes \\
\midrule
10-20 & ``teens'', ``10-20'' \\
20-30 & ``20-30'' \\
30-40 & ``30-40'' \\
40-50 & ``40-50'' \\
50-60 & ``50-60'' \\
\bottomrule
\end{tabular}
\end{table}

\begin{table}[H]
\centering
\small
\caption{Gaze Direction Harmonization}
\label{tab:harmonize_9}
\begin{tabular}{l p{0.6\textwidth}}
\toprule
Gaze Direction & Note Patterns \\
\midrule
Down & ``looking down'', ``looking at hands'', ``looking at the ground'', ``head in hands'', ``face in hands'', ``eyes closed'', ``looking at feet'', ``looking at lap'' \\
Up & ``looking up'', ``looking at the sky'', ``looking at sunset'', ``looking at sunrise'', ``looking at the cloud'', ``looking at clouds'', ``looking at lights'', ``looking to the sky'' \\
Out/Away & ``looking out'', ``looking outside'', ``looking at window'', ``out of window'', ``out window'', ``looking far'', ``looking afar'', ``looking into distance'', ``looking away'', ``looking around'', ``in front of window'', ``checking outside'' \\
\bottomrule
\end{tabular}
\end{table}

\section{Extended Quantitative Methods}

All feature extraction was implemented in Python using a custom analysis pipeline. The analysis code and video dataset analyzed in this study are available from the corresponding author upon reasonable request.

\subsection{Visual Aesthetics}

We extracted frames at one-second intervals, yielding 11 frames for 10-second app videos and 13 frames for 12-second API videos.

\subsubsection{Color and Light Measures}

Each measure was selected based on established connections to mood and emotion in prior research. Brightness, measured as perceptual luminance, has established connections to mood representation \cite{ref48}. Saturation captures color intensity; reduced saturation is associated with diminished emotional arousal \cite{ref49}. Colorfulness was computed using the Hasler-Süsstrunk metric \cite{ref50}, which correlates with human perception of chromatic vividness. Color temperature indexes the warm-cool balance; warm tones are associated with comfort and social connection, while cool tones connote isolation \cite{ref51}.

\begin{table}[H]
\centering
\small
\caption{Visual aesthetics measures}
\label{tab:methods_0}
\begin{tabular}{l p{0.65\textwidth}}
\toprule
Measure & Computation \\
\midrule
Brightness & $\text{Brightness} = 0.299R + 0.587G + 0.114B$ \\
Saturation & S channel of the HSV color space, computed as the difference between maximum and minimum RGB values divided by the maximum \\
Colorfulness & $\text{Colorfulness} = \sqrt{\sigma_{rg}^2 + \sigma_{yb}^2} + 0.3 \times \sqrt{\mu_{rg}^2 + \mu_{yb}^2}$ where $rg$ and $yb$ represent opponent color channels $(rg = R - G;\ yb = 0.5(R + G) - B)$, and $\sigma$ and $\mu$ denote standard deviation and mean respectively. \\
Color temperature & Mean $(R-B) + 0.3 \times (R-G)$ channel difference, with positive values indicating warmer colors and negative values indicating cooler colors. \\
Edge Density & Proportion of Canny edge pixels to total pixels: $\text{Edge Density} = \frac{\sum \mathbf{1}[\text{edge}(x,y) > 0]}{W \times H}$ where $W$ and $H$ are frame dimensions. Adaptive thresholds were set at 0.7$\times$ and 1.3$\times$ the median pixel intensity to account for varying image brightness. \\
\bottomrule
\end{tabular}
\end{table}

\subsubsection{Texture Measures}

We also extracted texture features using Gray-Level Co-occurrence Matrices \cite{ref30}, which provide insight into environmental qualities. Smooth textures may suggest sterile spaces, while complex textures indicate detailed or chaotic environments. We computed the GLCM at distance=1 pixel across four angles (0°, 45°, 90°, 135°) and averaged the resulting features.

where $P(i,j)$ is the normalized co-occurrence matrix.

\begin{table}[H]
\centering
\small
\caption{GLCM texture measures}
\label{tab:methods_1}
\begin{tabular}{l l p{0.35\textwidth}}
\toprule
Measure & Formula & Interpretation \\
\midrule
Contrast & $\sum_{i,j} (i-j)^2 \cdot P(i,j)$ & Local intensity variation \\
Dissimilarity & $\sum_{i,j} |i-j| \cdot P(i,j)$ & Average difference between neighboring pixels \\
Homogeneity & $\sum_{i,j} \frac{P(i,j)}{1 + (i-j)^2}$ & Closeness of distribution to diagonal \\
Energy & $\sum_{i,j} P(i,j)^2$ & Uniformity of gray level distribution \\
Correlation & $\sum_{i,j} \frac{(i - \mu_i)(j - \mu_j) \cdot P(i,j)}{\sigma_i \sigma_j}$ & Linear dependency of gray levels \\
Entropy & $-\sum_{i,j} P(i,j) \cdot \log_2(P(i,j))$ & Textural randomness or complexity \\
\bottomrule
\end{tabular}
\end{table}

\subsection{Audio Properties}

Audio was extracted from video files and analyzed using the librosa library \cite{ref31}. These acoustic features have established links to depressive affect: depressed speech is characterized by reduced volume, lower pitch, reduced pitch variability, and lower spectral centroid compared to non-depressed speech \cite{ref52,ref53}.

\begin{table}[H]
\centering
\small
\caption{Audio measures}
\label{tab:methods_2}
\begin{tabular}{l p{0.4\textwidth} p{0.3\textwidth}}
\toprule
Measure & Formula & Interpretation \\
\midrule
Volume & Root-mean-square (RMS) energy with frame length of 2048 samples and hop length of 512 samples & Acoustic intensity \\
Pitch & Fundamental frequency (F0) extracted using the pYIN algorithm \cite{ref54}, which combines a probabilistic YIN estimator with a hidden Markov model. Mean F0 computed across all voiced frames & Vocal register \\
Spectral centroid & $\text{Centroid} = \frac{\sum_{k} f_k \cdot M_k}{\sum_{k} M_k}$ where $f_k$ is the frequency of bin $k$ and $M_k$ is the magnitude. & ``Center of mass'' of the frequency spectrum; higher values indicate brighter, more energetic audio, lower values suggest muffled or subdued soundscapes \\
\bottomrule
\end{tabular}
\end{table}

\subsection{Semantic Content}

Spoken audio was transcribed using OpenAI's Whisper model (base) \cite{ref32}, with word-level timestamps extracted for temporal analysis. We computed sentiment scores, custom lexicons, and linguistic markers selected based on prior research linking these features to depression and recovery discourse.

\subsubsection{Sentiment and Speech Measures}

We utilized a VADER \cite{ref33} computed compound, positive, negative, and neutral sentiment ratings for each transcription. VADER is calibrated for social media text and handles informal language, emoticons, and emphasis markers.

\begin{table}[H]
\centering
\small
\caption{Semantic content measures}
\label{tab:methods_3}
\begin{tabular}{l p{0.65\textwidth}}
\toprule
Measure & Computation \\
\midrule
Word Count & Total words per video \\
Speech Rate & Words per second of video duration \\
Sentiment & VADER compound, positive, negative, and neutral scores \\
\bottomrule
\end{tabular}
\end{table}

\subsubsection{Light/Dark Lexicons}

We developed custom lexicons based on conceptual metaphor theory \cite{ref34} and research showing that depression discourse frequently employs metaphors of weight, darkness, and drowning, while recovery narratives invoke lightness and release \cite{ref55}. Using word-level timestamps from Whisper, we tracked when these terms appeared across the video timeline. Ambiguous terms that could appear in neutral contexts were excluded.

\begin{table}[H]
\centering
\small
\caption{Light/dark lexicons}
\label{tab:methods_4}
\begin{tabular}{l p{0.7\textwidth}}
\toprule
Lexicon & Terms \\
\midrule
Light (hope/recovery) & hope, hopeful, hoping, relief, relieved, release, released, free, freedom, peace, peaceful, calm, calming, serene, serenity, heal, healing, healed, recover, recovery, recovering, joy, joyful, happy, happiness, smile, smiling, laugh, laughing, laughter, love, loved, loving, comfort, comforting, comfortable, safe, safety, warm, warmth, support, supported, supporting, embrace, embraced \\
Dark (distress) & heavy, heaviness, burden, burdened, weigh, weighing, crush, crushing, crushed, drown, drowning, drowned, suffocate, suffocating, choke, choking, trap, trapped, stuck, confined, cage, caged, alone, lonely, loneliness, isolated, isolation, abandoned, forgotten, empty, emptiness, void, hollow, numb, numbness, pain, painful, hurt, hurting, ache, aching, suffer, suffering, exhaust, exhausted, exhausting, drain, drained, draining, weary, fatigue, despair, despairing, hopeless, hopelessness, helpless, helplessness, worthless, worthlessness, fear, afraid, scared, terrified, anxiety, anxious, panic, dread \\
\bottomrule
\end{tabular}
\end{table}

\subsubsection{Linguistic Pattern Measures}

We computed linguistic markers based on findings from computational studies of depression. First-person singular pronoun ratio indexes self-focused attention, which has been consistently linked to depressive rumination \cite{ref56}. Negation frequency captures negative cognitive framing characteristic of depressive thought patterns \cite{ref57}.

\begin{table}[H]
\centering
\small
\caption{Linguistic pattern measures}
\label{tab:methods_5}
\begin{tabular}{l p{0.40\textwidth} l}
\toprule
Measure & Numerator & Denominator \\
\midrule
First-person singular pronoun ratio & Count of: I, me, my, mine, myself & Total words \\
Negation ratio & Count of: no, not, never, nothing, none, nobody, nowhere, neither, nor, can't, couldn't, won't, wouldn't, don't, doesn't, didn't, shouldn't, isn't, aren't, wasn't, weren't, haven't, hasn't, hadn't & Total words \\
Lexical diversity (TTR) & Unique words & Total words \\
\bottomrule
\end{tabular}
\end{table}

\subsection{Temporal Dynamics}

We computed motion, scene structure, and feature trajectories to analyze whether videos depicted stasis, deterioration, recovery, or cyclical patterns over their duration. Motion magnitude is particularly relevant to depression representation because psychomotor retardation (slowed movement and activity) is a core diagnostic feature of major depression \cite{ref58}.

\subsubsection{Motion Analysis}

Dense optical flow was computed using the Farneb{\"a}ck algorithm \cite{ref35} between consecutive frames at native framerate (30 fps). The algorithm estimates motion vectors for every pixel using polynomial expansion. Flow magnitude was computed as $\sqrt{u^2 + v^2}$ where $u$ and $v$ are horizontal and vertical flow components. Frame-level values were binned into per-second averages for trajectory analysis.

\begin{table}[H]
\centering
\small
\caption{Farneb{\"a}ck optical flow parameters}
\label{tab:methods_6}
\begin{tabular}{ll}
\toprule
Parameter & Value \\
\midrule
Pyramid scale & 0.5 \\
Pyramid levels & 3 \\
Window size & 15 \\
Iterations & 3 \\
Polynomial neighborhood & 5 \\
Polynomial sigma & 1.2 \\
\bottomrule
\end{tabular}
\end{table}

\subsubsection{Scene Detection}

Scene cuts were detected by computing histogram correlation between frames sampled at one-second intervals. A cut was registered when correlation dropped below 0.3. Scene cut frequency provides insight into editing rhythm as rapid cuts may signal narrative progression or emotional intensity, while continuous shots suggest sustained states.

\subsubsection{Feature Trajectories}

For each video, per-second values were computed for all features. Linear slope was computed via ordinary least squares to characterize whether features increased, decreased, or remained stable over the video duration:
\begin{equation}
\text{slope} = \frac{\sum_{t}(t - \bar{t})(x_t - \bar{x})}{\sum_{t}(t - \bar{t})^2}
\end{equation}

\subsection{Statistical Formulas}

We used standard inferential statistics to compare features between App and API outputs, with corrections for multiple comparisons across all features tested.

\subsubsection{Group Comparisons}

Welch's t-test \cite{ref36} was used for all between-group comparisons because it does not assume equal variances across groups; an appropriate choice given the substantial differences in output characteristics between modalities.
\begin{equation}
t = \frac{\bar{X}_1 - \bar{X}_2}{\sqrt{\frac{s_1^2}{n_1} + \frac{s_2^2}{n_2}}}
\end{equation}

\subsubsection{Effect Sizes}

Cohen's d \cite{ref38} was used to quantify the magnitude of differences between App and API outputs, providing a standardized measure independent of sample size.
\begin{equation}
d = \frac{\bar{X}_1 - \bar{X}_2}{s_{\text{pooled}}}
\end{equation}
where
\begin{equation}
s_{\text{pooled}} = \sqrt{\frac{(n_1-1)s_1^2 + (n_2-1)s_2^2}{n_1+n_2-2}}
\end{equation}

Effect sizes were interpreted using conventional thresholds: $|d| = 0.2$ (small), $0.5$ (medium), $0.8$ (large).

\subsubsection{Multiple Comparison Correction}

To control the false discovery rate when conducting multiple comparisons, p-values were adjusted using the Benjamini-Hochberg procedure \cite{ref37} which provides a principled balance between Type I error control and statistical power:

The procedure operates as follows:

Rank all $m$ p-values from smallest to largest: $p_{(1)} \leq p_{(2)} \leq \ldots \leq p_{(m)}$

For each ranked p-value, compute the critical value: $\frac{k}{m} \times \alpha$

Find the largest $k$ such that $p_{(k)} \leq \frac{k}{m} \times \alpha$

Reject all null hypotheses for tests with rank $\leq k$.

Adjusted p-values (q-values) were computed as:
\begin{equation}
q_{(k)} = \min\left(\frac{m \cdot p_{(k)}}{k}, 1\right)
\end{equation}

The raw transformed values were then made monotone by processing from rank m down to rank 1, replacing each q(k) with the minimum of its own value and the q-value at rank k+1, ensuring that adjusted p-values respect the ordering of the original p-values. We used $\alpha$ = 0.05 for all analyses. Where relevant throughout the results, we report FDR-corrected q-values rather than raw p-values.

\subsection{Qualitative Interrater Reliability Methods}

Two authors (MF and ZZ) independently coded the complete dataset of 100 videos using the structured codebook, producing parallel sets of timestamped codes across all dimensions. One author's coding (MF) was designated as the primary dataset for analysis. The second author's independent coding (ZZ) was used to calculate inter-rater reliability.

\subsubsection{Cohen's Kappa}

Agreement on categorical dimensions was assessed using Cohen's kappa ($\kappa$) \cite{ref59}:
\begin{equation}
\kappa = \frac{p_o - p_e}{1 - p_e}
\end{equation}
where $p_o$ is the observed proportion of agreement and $p_e$ is the proportion of agreement expected by chance:
\begin{equation}
p_e = \sum_{k} p_{1k} \cdot p_{2k}
\end{equation}
where $p_{1k}$ and $p_{2k}$ are the proportions of each category assigned by coders 1 and 2, respectively.

Kappa was calculated for:
\begin{itemize}
\item Narrative shift presence (Yes/No)
\item Narrative direction (Recovery/Deterioration)
\item Environment categories (presence/absence per category per video)
\item Object categories (presence/absence per harmonized object per video)
\item Figure demographics (gender, race/ethnicity, age range, apparent SES)
\item Figure states (posture, facial expression, alone in frame)
\item Gaze direction (presence/absence of down, up, out/away per video)
\end{itemize}

Kappa values were interpreted following Landis and Koch \cite{ref60}:

\begin{table}[H]
\centering
\small
\caption{Landis and Koch kappa interpretation}
\label{tab:methods_7}
\begin{tabular}{ll}
\toprule
$\kappa$ Value & Interpretation \\
\midrule
< 0.00 & Poor \\
0.00--0.20 & Slight \\
0.21--0.40 & Fair \\
0.41--0.60 & Moderate \\
0.61--0.80 & Substantial \\
0.81--1.00 & Almost perfect \\
\bottomrule
\end{tabular}
\end{table}

\subsubsection{Temporal Intersection over Union}

For dimensions involving timestamped segments (environments, objects, figure states), kappa captures only whether coders agreed that a category was present in a video, not whether they agreed on when it appeared. To assess temporal agreement, we computed Intersection over Union (IoU) at one-second resolution:
\begin{equation}
\text{IoU} = \frac{|A \cap B|}{|A \cup B|}
\end{equation}
where A and B are binary temporal masks indicating the seconds during which each coder marked a category as present. For each video, we created binary vectors with one element per second of video duration (11 elements for 10-second app videos, 13 elements for 12-second API videos) for each coder's segment annotations, then computed the intersection and union across these vectors.

IoU ranges from 0 (no temporal overlap) to 1 (perfect temporal agreement). We computed IoU only for categories where both coders agreed the element was present, providing a measure of boundary precision conditional on presence agreement.

\subsubsection{Timestamp Precision}

For narrative shift timestamps (a single moment rather than a segment), we computed the mean and median absolute difference between coders' timestamps for videos where both identified a shift:
\begin{equation}
\text{Precision} = |t_{\text{MF}} - t_{\text{ZZ}}|
\end{equation}

Given the scale of the dataset (thousands of individual codes across six coding dimensions) item-by-item consensus resolution was not feasible. Reliability metrics were used to verify that the coding scheme could be applied consistently across coders, with the primary coder's judgments (MF) used for all analyses.

\section{Inter-Rater Reliability Results}

Two coders independently coded all 100 videos. The primary coder's data (MF) was used for all analyses; the second coder's independent coding (ZZ) was used to calculate reliability.

\subsection{Narrative Arc}

The three-way classification (recovery, deterioration, no shift) showed poor agreement on deterioration, so we collapsed to a binary classification of recovery vs. no recovery (no shift + deterioration), which achieved substantial agreement ($\kappa$ = 0.69, 85\% overall agreement). When both coders identified narrative shifts, temporal agreement was strong (mean difference = 0.31 seconds, median = 0.00 seconds, n = 35 pairs).

\subsection{Environments}

Overall presence/absence agreement was excellent ($\kappa$ = 0.91) with strong temporal agreement (mean IoU = 0.84). Category-specific kappas ranged from substantial to perfect:

\begin{table}[H]
\centering
\small
\caption{Environment inter-rater reliability}
\label{tab:reliability_0}
\begin{tabular}{lll}
\toprule
Category & $\kappa$ & Agreement \\
\midrule
Bathroom & 1.00 & 100\% \\
Kitchen/Dining & 1.00 & 100\% \\
Elevated Outdoor & 0.97 & 99\% \\
Urban Outdoor & 0.97 & 99\% \\
Underwater & 0.94 & 99\% \\
Transit & 0.91 & 98\% \\
Natural Outdoor & 0.89 & 97\% \\
Bedroom & 0.82 & 96\% \\
Other Interior & 0.78 & 94\% \\
Abstract & 0.66 & 98\% \\
At Window & 0.53 & 90\% \\
\bottomrule
\end{tabular}
\end{table}

\subsection{Objects}

Overall presence/absence kappa was moderate ($\kappa$ = 0.61) with strong temporal agreement when objects were mutually identified (mean IoU = 0.82). Agreement varied substantially by object type.

High reliability objects ($\kappa$ > 0.70): mug/cup ($\kappa$ = 0.93), pen ($\kappa$ = 0.90), mirror ($\kappa$ = 0.90), bench ($\kappa$ = 0.86), birds ($\kappa$ = 0.81), lamp ($\kappa$ = 0.81), cityscape/buildings ($\kappa$ = 0.80), cloud ($\kappa$ = 0.77), plant/flowers ($\kappa$ = 0.71).

Lower reliability objects: bed ($\kappa$ = 0.44), window ($\kappa$ = 0.47), sweatshirt/hoodie ($\kappa$ = 0.29), rain ($\kappa$ = 0.23). Examination of disagreements revealed an asymmetric pattern: the primary coder noted these ubiquitous items more exhaustively, while the secondary coder focused on more distinctive elements. This asymmetric pattern suggests the lower agreement values partly reflect a threshold difference in tagging exhaustiveness, though the resulting kappa values should be interpreted with this limitation in mind.

\subsection{Figure Demographics}

Gender achieved perfect agreement. Age showed high raw agreement (89\%) but low kappa ($\kappa$ = 0.38) due to restricted range (88\% of figures coded as 20-30). Race/ethnicity showed moderate agreement ($\kappa$ = 0.53), likely due to numerous bi-racial and ethnically ambiguous AI-generated figures; this dimension was excluded from analysis. Apparent SES showed poor agreement ($\kappa$ = 0.20) and was excluded from further analysis.

\begin{table}[H]
\centering
\small
\caption{Figure demographics inter-rater reliability}
\label{tab:reliability_1}
\begin{tabular}{lll}
\toprule
Variable & $\kappa$ & Agreement \\
\midrule
Gender & 1.00 & 100\% \\
Age & 0.38 & 89\% \\
Race/Ethnicity & 0.53 & 68\% \\
Apparent SES & 0.20 & 56\% \\
\bottomrule
\end{tabular}
\end{table}

\subsection{Figure States}

Posture and alone-in-frame status achieved strong reliability. Facial expression showed poor agreement ($\kappa$ = 0.49) due to ambiguous AI-generated expressions and was excluded from analysis.

\begin{table}[H]
\centering
\small
\caption{Temporal agreement (IoU)}
\label{tab:reliability_2}
\begin{tabular}{lll}
\toprule
Variable & $\kappa$ & IoU \\
\midrule
Posture & 0.89 & 0.86 \\
Alone in Frame & 0.94 & 0.98 \\
Facial Expression & 0.49 & 0.63 \\
\bottomrule
\end{tabular}
\end{table}

\subsection{Gaze Direction}

Coders frequently noted directional gaze in free-text annotations. When both coders documented gaze direction, agreement on dominant direction was high (96\%, n = 84). The primary coder noted downward gaze more frequently (93 videos) than the secondary coder (79 videos), reflecting threshold differences in annotation.

\subsection{Dimensions Excluded from Analysis}

Based on reliability results, the following dimensions were excluded from substantive analysis: race/ethnicity ($\kappa$ = 0.53), apparent SES ($\kappa$ = 0.20), and facial expression ($\kappa$ = 0.49).

\end{appendices}

\end{document}